\documentclass{SciPost}

\usepackage{microtype}          
\usepackage{amsmath}
\usepackage{amsfonts}
\usepackage{dsfont}
\usepackage{braket}
\usepackage{booktabs}
\usepackage{comment}
\usepackage{color}
\usepackage{xcolor}

\newcommand{\expval}[2][]{\mathbb{E}_{#1}\left[#2\right]}

\usepackage{listings}
\definecolor{codeframe}{HTML}{D0D4DE}
\usepackage[T1]{fontenc}
\usepackage[scaled=0.85]{beramono}
\definecolor{vsbg}{HTML}{FFFFFF}
\definecolor{vsfg}{HTML}{000000}
\definecolor{vsvar}{HTML}{001080}
\definecolor{vskeyword}{HTML}{0000FF}
\definecolor{vscontrol}{HTML}{AF00DB}
\definecolor{vsfunc}{HTML}{795E26}
\definecolor{vsclass}{HTML}{267F99}
\definecolor{vsstring}{HTML}{A31515}
\definecolor{vsnumber}{HTML}{098658}
\definecolor{vscomment}{HTML}{008000}
\definecolor{vslinenum}{HTML}{237893}
\lstdefinestyle{pythonstyle}{
    language=Python,
    backgroundcolor=\color{vsbg},
    basicstyle=\ttfamily\small\color{vsfg},
    identifierstyle=\color{vsvar},
    commentstyle=\color{vscomment},
    stringstyle=\color{vsstring},
    numbers=left,
    numberstyle=\tiny\color{vslinenum},
    frame=none,
    framesep=6pt, xleftmargin=12pt, framexleftmargin=12pt,
    showstringspaces=false,
    columns=fullflexible,
    keepspaces=true,
    upquote=true,
    breaklines=true,
    keywords=[1]{def, class, lambda, None, True, False, and, or, not, in, is,
                 global, nonlocal, self},
    keywordstyle=[1]\color{vskeyword},
    keywords=[2]{import, from, as, return, if, elif, else, for, while, try,
                 except, finally, with, yield, break, continue, pass, raise,
                 assert, del},
    keywordstyle=[2]\color{vscontrol},
    keywords=[3]{print, len, range, enumerate, zip, abs, sum, min, max},
    keywordstyle=[3]\color{vsfunc},
    keywords=[4]{int, float, str, list, dict, tuple,
                 SlaterJastrowBackflow, HOOrbitals, OrbitalJastrow,
                 ElectronCusp, OrbitalBackflow, NQS, RWM, MCSamplerCont},
    keywordstyle=[4]\color{vsclass},
    literate=*{0}{{\color{vsnumber}0}}1 {1}{{\color{vsnumber}1}}1
              {2}{{\color{vsnumber}2}}1 {3}{{\color{vsnumber}3}}1
              {4}{{\color{vsnumber}4}}1 {5}{{\color{vsnumber}5}}1
              {6}{{\color{vsnumber}6}}1 {7}{{\color{vsnumber}7}}1
              {8}{{\color{vsnumber}8}}1 {9}{{\color{vsnumber}9}}1,
}

\lstdefinestyle{outputstyle}{
    backgroundcolor=\color{white},
    basicstyle=\ttfamily\small,
    frame=single,
    rulecolor=\color{codeframe},
    framesep=6pt,
    xleftmargin=6pt, xrightmargin=6pt,
    breaklines=true,
    columns=fullflexible,
    mathescape=true
}

\hypersetup{
    colorlinks,
    linkcolor={red!50!black},
    citecolor={blue!50!black},
    urlcolor={blue!80!black}
}

\usepackage[bitstream-charter]{mathdesign}
\DeclareSymbolFont{usualmathcal}{OMS}{cmsy}{m}{n}
\DeclareSymbolFontAlphabet{\mathcal}{usualmathcal}

\fancypagestyle{SPstyle}{
\fancyhf{}
\lhead{\colorbox{scipostblue}{\bf \color{white} ~SciPost Physics Codebases }}
\rhead{{\bf \color{scipostdeepblue} ~Submission }}

\fancyfoot[C]{\textbf{\thepage}}
}

\begin{document}

\pagestyle{SPstyle}

\begin{center}{\Large \textbf{\color{scipostdeepblue}{
jVMC 2.0: A modular and scalable library for neural quantum states\\
}}}\end{center}

\begin{center}\textbf{
Massimo Solinas\textsuperscript{1$\star$}, Sven Benjamin Kožić\textsuperscript{1,2}, Wladislaw Krinitsin\textsuperscript{1,2}, Jonas B. Rigo\textsuperscript{1}, Noe Salmeron\textsuperscript{1} and Markus Schmitt\textsuperscript{1,2}
}\end{center}

\begin{center}
{\bf 1} University of Regensburg, 93053 Regensburg, Germany
\\
{\bf 2} Forschungszentrum J\"ulich GmbH, Peter Gr\"unberg Institute, Quantum Control (PGI-8), 52425 J\"ulich, Germany
\\[\baselineskip]
$\star$ \href{mailto:massimo.solinas@physik.uni-regensburg.de}{\small massimo.solinas@physik.uni-regensburg.de}
\end{center}

\section*{\color{scipostdeepblue}{Abstract}}
\textbf{\boldmath{%
We present jVMC 2.0, the second generation of the JAX-based jVMC library for variational Monte Carlo with neural quantum states. This new version was rebuilt from the ground up for large-scale simulations: samples, network evaluations and Jacobian products are sharded across multiple GPUs and nodes, and memory-efficient distributed linear algebra allows models with large numbers of parameters. A modular design around networks, samplers, operators, objective functions, solvers, time integrators and optimizers makes the library easy to extend, and it now supports continuous-space systems, fermions, symmetry projections, autoregressive and transformer architectures, and mixed-precision arithmetic. After benchmarking its scaling across devices, we showcase exemplary applications: ground states of the frustrated $J_1$--$J_2$ model with mixed precision and of electrons in a two-dimensional quantum dot, and quench dynamics in the two-dimensional transverse-field Ising model on lattices of up to $10\times10$ sites.
}}

\vspace{\baselineskip}

\noindent\textcolor{white!90!black}{%
\fbox{\parbox{0.975\linewidth}{%
\textcolor{white!40!black}{\begin{tabular}{lr}%
  \begin{minipage}{0.6\textwidth}%
    {\small Copyright attribution to authors. \newline
    This work is a submission to SciPost Physics Codebases. \newline
    License information to appear upon publication. \newline
    Publication information to appear upon publication.}
  \end{minipage} & \begin{minipage}{0.4\textwidth}
    {\small Received Date \newline Accepted Date \newline Published Date}%
  \end{minipage}
\end{tabular}}
}}
}


\vspace{10pt}
\noindent\rule{\textwidth}{1pt}
\tableofcontents
\noindent\rule{\textwidth}{1pt}
\vspace{10pt}


\section{Introduction}

The efficient simulation of quantum many-body systems remains one of the central challenges of computational physics, with implications ranging from condensed-matter physics and quantum chemistry to materials science and the development of quantum technologies. The fundamental obstacle is the exponential growth of the Hilbert space with the number of degrees of freedom: storing and manipulating the full wavefunction quickly becomes impossible, and full state methods remain limited to a few tens of spins even on the largest supercomputers~\cite{weisse2008exact, szabo2012modern}. Progress therefore relies on approximate methods that exploit the structure of physically relevant states, combined with algorithms that make efficient use of modern computing hardware. 

Over the past decades, the community has developed a variety of such methods, each with its own domain of applicability. Quantum Monte Carlo (QMC) techniques~\cite{foulkes2001quantum, becca2017quantum, assaad2008world} estimate observables by stochastic sampling and yield numerically exact results for large, unfrustrated bosonic and spin systems. For fermionic and frustrated systems, however, they suffer from the sign problem, whose cost grows exponentially with system size and inverse temperature~\cite{troyer2005computational}, and an analogous phase problem prevents their direct application to real-time dynamics. Tensor-network methods ~\cite{white1992density, schollwock2011density, orus2014practical}, do not suffer from the sign problem and provide quasi-exact results in one dimension, where low-energy states obey an area law of entanglement~\cite{eisert2010colloquium}. Their efficiency, however, relies on bounded entanglement so that their cost grows exponentially with the width of two-dimensional systems and with time in out-of-equilibrium dynamics. Dynamical mean-field theory (DMFT)~\cite{metzner1989correlated, georges1996dynamical} instead maps the lattice problem onto a self-consistent quantum impurity model and has become a powerful tool for strongly correlated materials~\cite{kotliar2006electronic}. Its central approximation, however, neglects non-local correlations, which can only be partially restored by cluster extensions at a substantially higher computational cost~\cite{maier2005quantum}.

A complementary route is offered by variational Monte Carlo (VMC)~\cite{becca2017quantum}, in which the wavefunction is represented by a parametrized ansatz and expectation values are estimated by stochastic sampling. VMC does not suffer from the sign problem and applies equally to bosons, fermions and frustrated systems, in any dimension and both in and out of equilibrium. Its accuracy, however, is limited by the bias of the chosen ansatz. 
%
%
%
Neural quantum states \cite{carleo2017solving} render VMC a systematically controlled numerical method by using artificial neural networks as variational ansätze~\cite{medvidovic2024neural, rigo2026neural, lange2024architectures}.
The universal approximation property of neural networks~\cite{hornik1991approximation} and their ability to efficiently encode large amounts of entanglement~\cite{deng2017quantum, levine2019quantum, sharir2022neural} make them compact yet highly expressive representations of the wavefunction, while their optimization directly benefits from GPU acceleration. NQS have been used both to find ground states, typically via stochastic reconfiguration~\cite{sorella1998green}, excited states~\cite{medvidovic2026adiabatic, pfau2024accurate} and to simulate unitary evolution through the time-dependent variational principle~\cite{carleo2017solving, schmitt2020quantum}. Over the years, the employed architectures have evolved from restricted Boltzmann machines~\cite{carleo2017solving} to convolutional~\cite{choo2019two}, autoregressive~\cite{sharir2020deep, hibat2020recurrent} and transformer-based networks~\cite{viteritti2023transformer}. The approach has proven successful for a broad range of systems, from frustrated spin models~\cite{choo2019two, viteritti2023transformer, chen2024empowering, wiersema2026numerical} and fermionic lattice models~\cite{chen2026neural, lange2026simulating} to electronic structure in continuous space~\cite{pfau2020ab, hermann2020deep, von2022self, fu2026towards}.

State-of-the-art NQS simulations now routinely employ networks with $10^5$--$10^6$ parameters and require a comparable number of samples per optimization step~\cite{chen2024empowering, rende2024simple}. At this scale, both the evaluation of the network on large batches of configurations and the solution of the dense linear systems arising in stochastic reconfiguration and the time-dependent variational principle easily exceed the memory and compute budget of a single GPU, making the distribution of the workload over multiple accelerators and compute nodes an essential requirement. At the same time, the development of efficient architectures and optimization schemes tailored to NQS remains an extremely active field of research~\cite{moss2025double, king2025beyond, duque2026one, krinitsin2026geometric}. Providing a common, well-tested and scalable code base that implements the core building blocks of the framework is therefore a fundamental task, both to ensure reproducibility and to lower the barrier for new methodological developments. 

With this work we continue this effort by presenting version 2.0 of jVMC~\cite{schmitt2021jvmc}, an open-source Python library built on JAX~\cite{jax2018github} and Flax~\cite{flax2020github} that implements the core components of variational Monte Carlo tailored to NQS. jVMC 2.0 retains the guiding philosophy of the first version, namely an easy-to-use and highly performant interface designed to be easily customized, modified and extended. However, the library has been almost completely rewritten: its components are now decoupled behind clean and uniform interfaces, so that each of them can be inspected, replaced or extended independently without touching the rest of the code. Beyond this restructuring, jVMC 2.0 introduces new functionalities including new architectures, new integrators, new samplers, compatibility with continuous VMC and new optimization tricks such as mixed precision schemes~\cite{solinas2026neural}. Most importantly, jVMC 2.0 provides a straightforward way of running simulations on multiple devices and compute nodes relying on JAX sharding mechanism. Sampling and network evaluations can be distributed across all available accelerators with no change to user code, while the dense linear algebra routines required by most of the optimization schemes can be now distributed over multiple GPUs through JAXMg~\cite{wiersema2026jaxmg}, removing the single-device memory bottleneck for large networks. \\

This paper is organized as follows. In Sec.~\ref{sec: vmc} we briefly review the variational Monte Carlo framework and its use for ground-state search and time evolution. Sec.~\ref{sec: codebase} describes the structure of the library and its main modules, and shows how users can build custom functionality on top of the existing components; Sec.~\ref{sec: multiple gpus} explains how simulations are distributed across multiple devices and nodes. With the codebase in place, we turn to applications: in Sec.~\ref{sec: benchmarks} we benchmark the performance of the library, and in Sec.~\ref{sec: examples} we apply it to representative problems in both ground-state search and unitary dynamics. We conclude in Sec.~\ref{sec: conclusion}; technical details on the gradient of the objective function and on memory-efficient distributed Jacobian products are given in the appendices.

\subsection{How to Install jVMC 2.0}
Since jVMC is built on JAX, it supports the same platforms, namely Linux and macOS, while GPU acceleration is only available on Linux. jVMC can be easily installed with
\begin{verbatim}
    pip install jVMC
\end{verbatim}
For GPU support, the CUDA version must be specified:
\begin{verbatim}
    pip install "jVMC[cuda12]"      # or "jVMC[cuda13]"
\end{verbatim}
This installs the corresponding CUDA-enabled build of JAX together with the CUDA libraries, as well as JAXMg for distributed linear algebra across multiple GPUs.

On computing clusters, where CUDA is usually provided through environment modules, we recommend using the system installation instead. After loading the relevant modules, jVMC can be installed with the \texttt{-local} variant:
\begin{verbatim}
    module load cuda                   # cluster-specific
    pip install "jVMC[cuda12-local]"   # or "jVMC[cuda13-local]"
\end{verbatim}
In this way JAX uses the CUDA libraries that the cluster provides, matched to the local drivers. This avoids version conflicts with a second, pip-installed copy of CUDA.

\subsection{Migrating from jVMC 1.x and using AI coding agents}
\label{sec: migration}

Since jVMC~2.0 is a near-complete rewrite, scripts written for version~1.x do not run without changes: the interface now follows PEP~8 naming, MPI has been replaced by JAX sharding, and components such as solvers, steppers and objective functions are now separate objects. To ease the transition, the repository contains a migration guide (\texttt{MIGRATION.md}, also part of
the online documentation). It shows a version-1.x script next to its version-2.0 equivalent, maps the old names to the new ones module by module, and lists the changes in behavior that do
not raise errors. \\

AI coding assistants are increasingly used to write simulation code. Since such models are trained on publicly available code, they are likely to reproduce the interface of jVMC~1.x. We
therefore distribute, in the folder \texttt{skills/jvmc} of the repository, an \emph{agent skill}: a set of Markdown files in the open Agent Skills format, which compatible agents load when a task involves the library. The skill describes the structure
of a jVMC calculation, gives complete examples of ground-state searches and real-time evolution, lists common pitfalls and maps the version-1.x names to the current ones. 

\subsection{The Variational Monte Carlo Framework}\label{sec: vmc}

\begin{figure}[t]
    \centering
    \includegraphics[width=\linewidth]{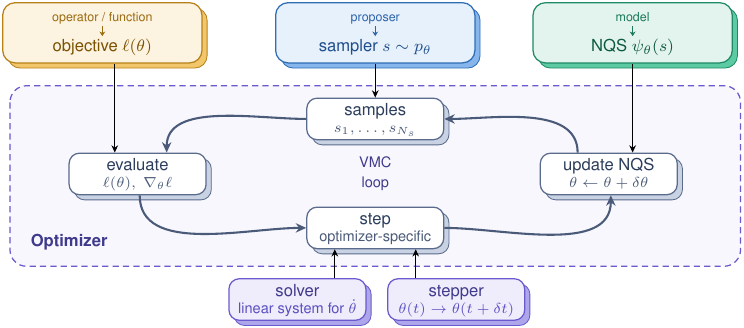}
    \caption{Schematic of the VMC optimization loop in jVMC. The user specifies three
    components (top): an objective $\ell(\theta)$, built from an operator or a generic
    function; a sampler, which draws configurations $s \sim p_\theta$ using a chosen
    proposer; and the model, a neural quantum state $\psi_\theta(s)$. Each iteration of the
    loop, handled by the optimizer (dashed region), proceeds as follows. A set of samples
    $s_1, \ldots, s_{N_s}$ is drawn from the current model and used to estimate the objective
    $\ell(\theta)$ and its gradient $\nabla_\theta \ell$. An optimizer-specific step then
    computes the parameter update $\delta\theta$, and the network is updated as
    $\theta \leftarrow \theta + \delta\theta$ before new samples are drawn. For
    natural-gradient and time-dependent schemes, the step relies on two further
    interchangeable components (bottom): a solver for the linear system determining
    $\dot\theta$, and a stepper that integrates $\theta(t) \rightarrow \theta(t + \delta t)$.}
    \label{fig: vmc scheme}
\end{figure}

NQS build on the well-established VMC framework. Its starting point is the observation that the quantum expectation value of an observable $\hat O$ can be rewritten as a classical expectation value over the Born
distribution in a given basis $\left\{\ket{s}\right\}$:
\begin{equation}\label{eq: expval}
    \frac{\bra{\psi_\theta}\hat O\ket{\psi_\theta}}{\bra{\psi_\theta}\psi_\theta\rangle}=\expval[p_\theta]{\frac{\left(\hat O\psi_\theta\right)(s)}{\psi_\theta(s)}}:=\expval[p_\theta]{O_{\text{loc}}(s)}, \quad p_\theta(s)=\frac{|\psi_\theta(s)|^2}{\mathcal{N}},
\end{equation}
where $\mathcal{N}$ represents the normalization of the distribution, and we defined the local observables $O_{\text{loc}}$.

With this basis, VMC rests on three ingredients:
(i) a compact representation $\psi_\theta(s) = \bra{s}\psi_\theta\rangle$ of the wave
function amplitudes, depending on a set of variational parameters $\theta$ whose number
scales polynomially with system size, which circumvents the exponential size of the
Hilbert space;
(ii) Monte Carlo sampling from $p_\theta$, which replaces Eq.~\eqref{eq: expval} by a sample
mean over $N_s$ configurations, so that the wavefunction only has to be evaluated on typical configurations that appear in the sample; (iii) a variational principle that prescribes how $\theta$ is updated to approach a target
state, such as the ground state, an excited state, or the time-evolved state under unitary dynamics.

What distinguishes NQS from earlier VMC approaches is the choice of the ansatz. Traditional VMC relies on physically motivated wavefunctions, such as Jastrow or Gutzwiller-projected states, with few parameters. These encode strong prior knowledge and are cheap to optimize, but they introduce a systematic bias whenever that prior does not capture the physics of the target state. NQS replace them with generic, highly expressive artificial neural network. This reduces the systematic bias of the ansatz at the cost of a typically much larger parameter space and a more demanding stochastic optimization, which becomes tractable by borrowing architectures, automatic differentiation, and hardware acceleration from modern machine learning.

From this perspective, VMC with NQS can be viewed as an unsupervised learning problem in
which the training data, the configurations $s$, are not given in advance but are generated
at every iteration by the model itself. Once a network (the \emph{model}), a procedure for
generating configurations (the \emph{sampler}), and an objective function are specified, VMC
reduces to an optimization loop: configurations are sampled, used to estimate the value and
the gradient of the objective, and these estimates are used to update the variational
parameters. This loop is shown schematically in Fig.~\ref{fig: vmc scheme}.
\section{Codebase Design and Features} \label{sec: codebase}
As outlined in the introduction, a VMC codebase rests on three core ingredients: a network architecture that represents the wavefunction, a procedure for drawing configurations from the associated probability distribution, and a means of defining operators and evaluating their quantum expectation values. Building on these three components, one can implement different algorithms that drive the optimization of the network towards the desired wavefunction, what we refer to as the optimizer. \\
This structure translates in code in defining a series of core modules, that together define an optimization task, as illustrated in this simple example:

\begin{lstlisting}[style=pythonstyle]
import jVMC

# Problem and numerical hyperparameters
L = 10                  # number of sites
n_samples = 2**12       # samples drawn at every step
n_chains = 2**10        # parallel Markov chains
batch_size = n_samples  # configurations evaluated at once
steps = 100             # number of optimization steps
learning_rate = 1e-3    # time step of the integrator

# 1. Network: log-amplitude of a single configuration
net = jVMC.nets.RBM()

# 2. NQS: vectorized evaluation of the network and its gradients
psi = jVMC.vqs.NQS(net, L, batch_size)

# 3. Sampler: the way to generate new samples
proposer = jVMC.propose.SpinFlip()
sampler = jVMC.sampler.MCSampler(
    psi, proposer, num_chains=n_chains, num_samples=n_samples
)

# 4. Operator: observable to be used in the objective function
H = sum(jVMC.operator.discrete.SigmaX(i) for i in range(L))

# 5. Objective function: expectation value of the operator
loss = jVMC.objective_function.Observable(H)

# 6. Stepper: time integrator for the parameters
stepper = jVMC.stepper.Euler(learning_rate)

# 7. Solver: linear system solver used in the update
solver = jVMC.solver.Pinv()

# 8. Optimizer: orchestrates the VMC loop
opt = jVMC.optimizer.MinSR(sampler, psi, solver=solver)

# Run the optimization
out = opt.ground_state_search(steps, loss, stepper)
\end{lstlisting}

In what follows, we describe each of these building blocks of our codebase in detail, outlining both how they are implemented and, more importantly, how they can be extended to develop and experiment with new custom methods.

\subsection{Defining the Neural Quantum State}
The central object of jVMC is the neural quantum state, which provides the
variational ansatz for VMC. For a given configuration $s$, it encodes the
logarithmic amplitude $\log\langle s|\psi_\theta\rangle$, where $\theta$ denotes the set of variational parameters. Note that working in log-space is the standard choice in VMC frameworks, since unnormalized amplitudes can vary over many orders of magnitude and would otherwise lead to numerical instabilities during optimization. The first step for the user is therefore to define the underlying network, which we refer to as \texttt{NET}, that maps a configuration $s$ to $\log\langle s|\psi_\theta\rangle$.

\subsubsection{Network} \label{sec: net}
Our library relies on Flax~\cite{flax2020github}, which allows neural networks to be defined in a straightforward way on top of JAX~\cite{jax2018github}. Any custom network can be implemented using either the Flax \texttt{nn} or \texttt{nnx} API. Since Flax is gradually transitioning from \texttt{nn} to \texttt{nnx}, the \texttt{NQS} class is compatible with networks defined in either framework: under the hood, an \texttt{nnx} model is automatically converted into an equivalent \texttt{nn} module. The only requirement imposed on the user is that the network's \texttt{\_\_call\_\_} method be defined for a single input configuration of shape \texttt{(sample\_shape,)}; the \texttt{NQS} class then automatically vectorizes this computation over arbitrary batches of configurations using \texttt{jax.vmap}. This design choice keeps the definition of new network architectures as simple as possible.

In addition, the \texttt{NET} can optionally implement two further methods, \texttt{sample} and \texttt{eval\_ratio}. The former is discussed in Sec.~\ref{sec: autoregressive}, while the latter defines a function that, given two configurations $s$ and $s'$, directly returns their amplitude ratio,
\begin{equation}
    \texttt{NET.eval\_ratio}: (s, s') \longmapsto \frac{\psi_\theta(s')}{\psi_\theta(s)}.
\end{equation}
This functionality is useful whenever an analytical expression for the ratio is available, since evaluating it directly is computationally cheaper than calling the model on the two configurations separately. \\

When the user wishes to represent the real and imaginary parts of the amplitude with two separate networks, these can be passed to the \texttt{TwoNets} wrapper, available as \texttt{jVMC.nets.two\_nets\_wrapper}, before being handed to the \texttt{NQS} class. This wrapper guarantees that the combined output is consistently complex-valued, and that extracting the real part of the amplitude only triggers evaluation of the network responsible for it.

Finally, jVMC provides a collection of pre-implemented network architectures (see \texttt{jVMC.nets}) that users can readily employ in their VMC calculations. These range from restricted Boltzmann machines~\cite{carleo2017solving} and convolutional networks~\cite{choo2019two} to autoregressive models, namely recurrent networks with GRU/LSTM cells in one and two dimensions~\cite{hibat2020recurrent} and RWKV-type architectures~\cite{peng2023rwkv}, which allow exact sampling, as well as Vision Transformers~\cite{viteritti2023transformer} and a hidden-fermion Pfaffian extension for fermionic systems. All networks are standard Flax modules, so they can be freely combined with or replaced by user-defined architectures.

\subsubsection{NQS} \label{sec: nqs}
Once the network has been defined, it can be handed to the \texttt{NQS} class,
which wraps it and exposes a simple interface to inspect and overwrite the
variational parameters $\theta$, either in their native dictionary/pytree
structure, as used internally by Flax (\texttt{NQS.parameters}), or as a single
flat array (\texttt{NQS.parameters\_flat}). Since the network can be defined
with either real or complex parameters, the \texttt{NQS} class adopts the
convention that the flat representation is always a real-valued array. For a
network with $N_p$ complex parameters, this array has length $2N_p$ and
collects their real and imaginary parts,
\begin{equation}
    \texttt{NQS.parameters\_flat} = \left(\mathrm{Re}\,\theta,\ \mathrm{Im}\,\theta\right) \in \mathbb{R}^{2N_p},
\end{equation}
while for real parameters it simply coincides with $\theta \in \mathbb{R}^{N_p}$.
In other words, the \texttt{NQS} class always treats the network as a function of
real variational parameters, and complex parameters are handled by doubling
their number.

The class is responsible for the following functionalities: 
\begin{itemize}
    \item \textbf{Defining the Batch Size}: During initialization, the user specifies the \texttt{batch\_size}, i.e., the number of input configurations that the \texttt{NQS} processes simultaneously when evaluating the network or its gradients. This parameter is essential to control the tradeoff between execution time and memory, and has to be tuned looking at the number of samples produced by the \texttt{Sampler} class (see Sec.~\ref{sec: sampler}). Setting \texttt{batch\_size} equal to the total number of samples $N_s$ maximizes throughput, since all evaluations are then performed in a single vectorized call. On the other hand, reducing it trades some speed for a smaller memory footprint, which becomes necessary whenever the full batch of network evaluations and gradients would otherwise exceed the available memory (see Sec.~\ref{sec: batching vs memory}).
    \item \textbf{Evaluation}: Given a set of configurations $S=\{s_i\}_{i=1}^{N_s}$, the class returns the log-amplitude of the configurations,
    \begin{equation}
        \texttt{NQS}.\_\_\texttt{call}\_\_: S \longmapsto \log\psi_\theta(S)\in I,
    \end{equation}
    where $I$ denotes either $\mathbb{R}$ or $\mathbb{C}$, depending on the chosen parametrization. The user can decide to define a net whose output can represent either the log-amplitude directly or the bare amplitude. In the latter case, the user must indicate this when initializing the \texttt{NQS} class by setting the flag \texttt{logarithmic} to \texttt{False}, so that the logarithm is then applied internally.
    \item \textbf{Gradient}: Given a set of configurations $S=\{s_i\}_{i=1}^{N_s}$,
    the class computes the Jacobian matrix of logarithmic derivatives $\Gamma$, with
    elements
    \begin{equation}\label{eq: Jacobian def}
        \Gamma_{i,k}=\Gamma_k(s_i)=\frac{\partial\log\psi_\theta(s_i)}{\partial\theta_k},
    \end{equation}
    through the method
    \begin{equation}
        \texttt{NQS.gradients}: S \longmapsto \Gamma .
    \end{equation}
    Consistently with the parameter convention above, the derivatives are
    always taken with respect to the real parameters stored in
    \texttt{NQS.parameters\_flat}. Hence $\Gamma$ has shape $N_s\times N_p$ for
    real parameters and $N_s\times 2N_p$ for complex parameters $\theta_k = a_k + i b_k$,
    in which case
    \begin{equation}\label{eq: complex gamma}
        \Gamma(s) = \left(\frac{\partial\log\psi_\theta(s)}{\partial a},
                         \frac{\partial\log\psi_\theta(s)}{\partial b}\right)=\left(\Gamma^{(a)},\Gamma^{(b)}\right).
    \end{equation}
    Note that $\Gamma$ is in general complex even when $\theta$ is real, since
    $\log\psi_\theta$ is complex-valued. 
    How $\Gamma$ is evaluated depends on the analytic structure of the network,
    which the \texttt{NQS} class determines automatically:
    \begin{itemize}
        \item \emph{Real parameters.} If $\log\psi_\theta$ is real, $\Gamma$ is
        obtained by a single automatic differentiation. Otherwise, the real
        and imaginary parts of $\log\psi_\theta$ are differentiated separately
        and recombined as
        $\Gamma_k = \partial_{\theta_k}\mathrm{Re}\log\psi_\theta + i\,\partial_{\theta_k}\mathrm{Im}\log\psi_\theta$.

        \item \emph{Complex parameters, non-holomorphic network.} The same
        procedure is applied, now differentiating with respect to both $a_k$
        and $b_k$, yielding the $2N_p$ columns of $\Gamma$.

        \item \emph{Complex parameters, holomorphic network.} If
        $\log\psi_\theta$ is holomorphic in $\theta$, the Cauchy--Riemann
        conditions imply $\Gamma^{(b)}=i\Gamma^{(a)}$, so that the Jacobian follows from a single differentiation of the real part of $\log\psi_\theta$.
    \end{itemize}
    Holomorphicity is probed numerically on a sample input configuration by gauging the validity of the Cauchy--Riemann conditions.
    \item \textbf{Lazy gradient}: Since materializing the full Jacobian $\Gamma$ becomes memory-intensive for large $N_s$ or $N_p$, the \texttt{NQS} class also provides the \texttt{lazy\_gradients} method. Instead of computing $\Gamma$ at once, it returns a reusable generator that yields it in consecutive blocks of \texttt{batch\_size} rows, so that only an array of shape $\texttt{batch\_size}\times N_p$ is held in memory at any time. This is essential for objective functions that accumulate statistics over many configurations (see Sec.~\ref{sec: objective function}).
    
    \item \textbf{Mixed Precision}: the \texttt{NQS} class can be provided with three additional parameters at initialization, \texttt{sampler\_dtype}, \texttt{eval\_dtype}, and \texttt{grad\_dtype}, which together enable mixed-precision VMC. As their names suggest, \texttt{sampler\_dtype} controls the dtype used for the network parameters (and hence for all internal operations) when the network is called for sampling purposes, either directly by the \texttt{NQS} or by the \texttt{Sampler}; \texttt{eval\_dtype} controls the dtype used when evaluating the network via \texttt{NQS.\_\_call\_\_}; and \texttt{grad\_dtype} controls the dtype used when computing gradients via \texttt{NQS.gradients}. In all three cases, the output of the network is subsequently cast to a fixed output dtype, given by the global variable \texttt{DT\_OUT\_REAL} or \texttt{DT\_OUT\_CPX}, depending on whether the network output is real- or complex-valued. This design decouples the precision used internally, in the computationally expensive network evaluations, from the precision used to accumulate and report statistics, giving the user fine-grained control over a mixed-precision optimization scheme. Section~\ref{sec:mixed_precision} presents a ground-state search using mixed precision. If the user does not specify these three parameters explicitly, they default to the global variable \texttt{DT\_PARAMS\_REAL} (or \texttt{DT\_PARAMS\_CPX}, for complex-valued parameters). For this mechanism to function correctly, the network must be defined so that it can accept parameters of varying dtype. In Flax, this is achieved simply by not hard-coding a \texttt{dtype} argument in the network's layers: specifying a fixed \texttt{dtype} forces all computations inside that layer to that dtype, whereas omitting it lets JAX infer the dtype directly from the parameters passed at call time, which is what allows the \texttt{NQS} class to switch precision on the fly.
\end{itemize}

\subsubsection{Symmetrizing the Network}\label{sec: sym net}

Symmetries of the Hamiltonian can be imposed on an NQS either through the network architecture or by projecting its amplitudes onto a symmetry sector.
In the first approach, the internal features transform equivariantly under the chosen symmetry, and an invariant output is obtained by pooling these features.
For example, the \texttt{ViterittiViT} architecture uses translation-invariant attention and a sum over patch features to make its backbone invariant under translations by complete patches~\cite{rende2024simple}.
An invariant scalar output belongs to the trivial symmetry sector, whereas a nontrivial character requires an appropriately transforming output or an additional projection~\cite{vicentini2021netket}.

For a finite symmetry group $G$ with unitary operators $\hat T_g$ and characters $\chi_g$, the symmetry projector is
\begin{equation}\label{eq: symmetry projection}
    \begin{aligned}
        \hat P_\chi &= \frac{d_{\chi}}{|G|}\sum_{g\in G}\chi_g^*\hat T_g,
    \end{aligned}
\end{equation}
which produces an unnormalized projected state with wavefunction amplitudes
\begin{equation}\label{eq: symmetry projection}
    \begin{aligned}
        \ket{\psi_{\theta,\chi}} &= \hat P_\chi\ket{\psi_\theta}
        = \frac{d_{\chi}}{|G|}\sum_{g\in G}\chi_g^*\hat T_g\ket{\psi_\theta},\\
        \psi_{\theta,\chi}(s) &= \langle s|\psi_{\theta,\chi}\rangle
        = \frac{d_{\chi}}{|G|}\sum_{g\in G}\chi_g^*\eta_g(s)\psi_\theta(g^{-1}s),
    \end{aligned}
\end{equation}
where $\eta_g(s)=\langle s|\hat T_g|g^{-1}s\rangle$ includes any phase or sign associated with the transformation~\cite{vicentini2021netket}.
For permutations of spin or bosonic occupations, this factor is unity and the permutation sign of the state $g^{-1}s$ for fermions. 
The \texttt{SymmetryProjector} class represents the transformed configurations by permutations of site indices, optionally combined with local occupation transformations, and stores the scalar coefficients $\chi_g^*$ in its \texttt{character} field.
The default averaging mode, \texttt{symmetry\_average="exp"}, evaluates the logarithm of the weighted amplitude sum in Eq.~\eqref{eq: symmetry projection} using a numerically stabilized exponential average.
Thus, the projection acts on amplitudes even though the underlying network returns their logarithms.
This construction also applies to one-dimensional sectors of non-Abelian groups, although only Abelian symmetries are currently available.

The lattice constructors provide periodic translations with a specified lattice momentum, the square point group $D_4$, the rectangular point group $D_2$, reflection of a chain, global spin inversion, and particle--hole transformations for the supported particle types.
Spatial permutations can use the built-in direct or snake site ordering, or a user-supplied site enumeration.
The point-group constructors select the fully symmetric sector by default.
The overloaded multiplication applies a projector to a network according to
\begin{equation}\label{eq: symmetry network interface}
    \texttt{*}: (\hat P_\chi,\,\texttt{NET}) \longmapsto \texttt{NET}_\chi,
    \qquad \texttt{NET}_\chi(s)=\log\psi_{\theta,\chi}(s).
\end{equation}
The returned network is wrapped by \texttt{ProjectedOrbitNet} and can be passed to \texttt{NQS}.
The same operation composes two projectors,
\begin{equation}\label{eq: symmetry composition interface}
    \texttt{*}: (\hat P_1,\,\hat P_2) \longmapsto \hat P_1\hat P_2.
\end{equation}
This product represents simultaneous projection onto the two sectors when the projectors commute.
For example, zero momentum is compatible with the full square point group, giving
\begin{lstlisting}[style=pythonstyle]
import jVMC
from jVMC.symmetry import lattice_symmetries as sym

Lx = Ly = 4
raw_net = jVMC.nets.RBM(num_hidden=4, bias=True)
D4 = sym.square_d4_symmetry(Lx, Ly, particle_type="spin")
T_square = sym.square_translation_symmetry(
    Lx, Ly, particle_type="spin", q=(0, 0)
)
net = (D4 * T_square) * raw_net
psi = jVMC.vqs.NQS(net, Lx * Ly, batch_size=16, seed=7)
\end{lstlisting}

Custom projectors are defined by supplying the index permutations and scalar weights for the desired sector, together with any required local occupation transformations.
As an example, consider spins on the eight vertices of a cube.
Quarter-turn rotations about one of its three fourfold (tetrad) axes generate the group $C_4=\{e,R,R^2,R^3\}$.
Number the vertices cyclically as $0,1,2,3$ on one face and $4,5,6,7$ on the opposite face, with the same orientation about the chosen axis.
The nontrivial sector with $\chi(R)=i$ is selected by the conjugate character weights $(1,-i,-1,i)$:
\begin{lstlisting}[style=pythonstyle]
import numpy as np

cube_faces = np.arange(8).reshape(2, 4)
c4_permutations = tuple(
    tuple(np.roll(cube_faces, -r, axis=1).ravel())
    for r in range(4)
)
C4 = jVMC.symmetry.SymmetryProjector(
    Lx=4, Ly=2, particle_type="spin",
    perm=c4_permutations, character=(1, -1j, -1, 1j),
    name="cube_C4"
)
net_c4 = C4 * raw_net
\end{lstlisting}
Here, \texttt{Lx=4} and \texttt{Ly=2} enumerate four vertices on each of two faces.
The permutations specify the rotation of the cube, each row of \texttt{perm} produces $R^{-r}s$ for evaluating Eq.~\eqref{eq: symmetry projection}.
For a nonzero projection, the resulting state has eigenvalue $i$ under the quarter-turn operator $\hat R$.
More generally, the supplied operations and weights must form a consistent representation of the chosen symmetry group.

For fermions, permuting the occupations also permutes the creation operators defining the Fock basis, and the resulting occupation-dependent parity must be included in $\eta_g(s)$.
Selecting \texttt{particle\_type="spinless\_fermion"} or \texttt{"spinful\_fermion"} in the constructors enables these signs in the library's fixed Fock ordering.
Spinful configurations contain one integer per site, with $0,1,2,3$ denoting empty, spin-up, spin-down, and double occupation, respectively.
The implementation also includes the local fermionic phases required by spin inversion and particle--hole transformations.

Besides coherent amplitude projection, \texttt{ProjectedOrbitNet} provides two nonlinear averages: \texttt{"log"} averages logarithmic amplitudes directly, while \texttt{"sep"} averages the probabilities and the phases separately, using a circular mean for the latter.
Both modes ignore the scalar projector weights and fermionic signs, so they do not implement Eq.~\eqref{eq: symmetry projection} for nontrivial characters or sign factors.
For networks evaluated with MCMC, \texttt{"exp"} is therefore the appropriate choice for such sectors.
The average can also be supplied as a callable taking the logarithmic amplitudes and their weights and returning one logarithmic amplitude.
Table~\ref{tab: symmetry averages} summarizes the symmetry and network restrictions of these choices.

\begin{table}[htbp]
    \centering
    \caption{Averaging choices in \texttt{ProjectedOrbitNet}.}
    \label{tab: symmetry averages}
    \small
    \setlength{\tabcolsep}{4pt}
    \begin{tabular}{@{}p{0.13\linewidth}p{0.405\linewidth}p{0.405\linewidth}@{}}
        \toprule
        \textbf{Average} & \textbf{Symmetry restriction} & \textbf{Network restriction} \\
        \midrule
        \texttt{"exp"} & All supported characters and sign factors. & No direct-sampling interface. \\
        \addlinespace
        \texttt{"log"} & Unit weights for the intended symmetry sector; all supplied weights are ignored. & No direct-sampling interface. \\
        \addlinespace
        \texttt{"sep"} & Unit weights for the intended symmetry sector; all supplied weights are ignored. & With or without direct sampling; direct sampling assumes \texttt{log\_prob\_factor} $=0.5$. \\
        \addlinespace
        \emph{callable} & Determined by the callable; direct-sampling networks still require unit weights. & With or without direct sampling; for direct sampling, the squared modulus of the amplitude must reproduce the wrapper's probability mixture. \\
        \bottomrule
    \end{tabular}
\end{table}

If the underlying network provides autoregressive sampling, the wrapper accepts only \texttt{"sep"} or a custom callable, and rejects nontrivial characters and additional phase or sign factors.
The built-in \texttt{"sep"} mode assumes the pure-state convention \texttt{log\_prob\_factor=0.5} and preserves direct sampling by combining a sample from the original network with a uniformly chosen symmetry transformation~\cite{hibat2020recurrent}.
The resulting probability is the average of the transformed Born probabilities; coherent projection generally produces a different distribution.
A custom average must preserve consistency with this sampling rule.

\subsection{Sampling Configurations} \label{sec: sampler}

The second core ingredient of the VMC framework is the \texttt{Sampler}, an object responsible for generating configurations to be fed to the \texttt{NQS} and for computing expectation values of user-specified observables. These two responsibilities are captured by the \texttt{AbstractSampler} base class, which requires only an \texttt{NQS} instance for initialization. The class stores the most recently generated configurations $S$ (the samples), together with their log-amplitudes and their associated probabilities (the weights). Since the class already holds all the ingredients required to compute quantum expectation values, its call method can be used directly for this purpose,
\begin{equation}
    \texttt{AbstractSampler.\_\_call\_\_}: \hat O \longmapsto \frac{\langle\psi_\theta|\hat O|\psi_\theta\rangle}{\langle\psi_\theta|\psi_\theta\rangle}.
\end{equation}
The result is returned as a \texttt{SampledObs} object (see Sec.~\ref{sec: SampledObs}).

From this general interface, two different kinds of samplers are provided to the user: the \texttt{ExactSampler} and the \texttt{AbstractMCSampler}.

\subsubsection{Exact Sampler}

The \texttt{ExactSampler} is designed to enumerate and generate the complete basis of a discrete Hilbert space $\mathcal{H}_d$. It is initialized by providing the local Hilbert space dimension \texttt{l\_dim} and the parameter \texttt{log\_prob\_factor}, which controls the normalization convention used for the returned probabilities, as detailed below. The class infers the total number of sites $N$ from the sample shape and efficiently constructs the basis,
\begin{equation}
    S = \left[s_1, s_2, \ldots, s_M\right], \qquad M = \texttt{l\_dim}^{N}.
\end{equation}
This is done through the cached property \texttt{ExactSampler.basis}. The \texttt{sample} method then simply returns these configurations together with their amplitudes and probabilities,
\begin{equation}
    \texttt{ExactSampler.sample}: \; \longmapsto \; \left(S,\; \log\psi_\theta(S),\; \frac{|\psi_\theta(S)|^{1/\texttt{log\_prob\_factor}}}{\mathcal{N}}\right),
\end{equation}
where $\mathcal{N} = \sum_{s \in S} |\psi_\theta(s)|^{\frac{1}{\texttt{log\_prob\_factor}}}$ is a normalization constant ensuring that the returned probabilities sum to one.

\subsubsection{MC Sampler}

The \texttt{AbstractMCSampler} implements a more elaborate interface, whose purpose is to draw configurations from the probability distribution
\begin{equation}
    p_{\mu,\theta}(s) = \frac{|\psi_\theta(s)|^{\mu}}{\sum_s |\psi_\theta(s)|^{\mu}}, \qquad \mu \in (0, 2].
\end{equation}
This is achieved by means of a Markov Chain Monte Carlo (MCMC) or auto-regressive sampling (which we discuss in Sec.~\ref{sec: autoregressive}) if the network itself can generate samples. In the former case, given a configuration $s$, a new configuration $s'$ is proposed and accepted with the Metropolis--Hastings acceptance probability,
\begin{equation}\label{eq: acceptance ratio}
    \alpha(s,s') = \min\left(1, \frac{p_{\mu,\theta}(s')}{p_{\mu,\theta}(s)}\, C(s, s')\right),
\end{equation}
where $C(s,s') = T(s' \to s)/T(s \to s')$ denotes the proposal ratio, i.e., the probability of proposing the reverse transition from $s'$ to $s$ divided by the probability of proposing the forward transition from $s$ to $s'$, under whichever proposal distribution has been chosen by the user (see Sec.~\ref{sec: proposer}).

Accordingly, when initializing the \texttt{AbstractMCSampler}, the user must specify the desired number of samples and the number of independent Markov chains to run. Generation of the independent chains is efficiently parallelized; if the code is run on multiple devices, this parallelization is automatically distributed across them (see Sec.~\ref{sec: multiple gpus}). Internally, the class adjusts the requested numbers of samples and chains so that the number of chains is always divisible by the number of available devices.

To ensure proper decorrelation between successive samples, the user can tune the number of sweep steps, i.e., the number of Monte Carlo proposal/acceptance moves performed between two configurations that are actually retained as samples. Discarding these intermediate configurations reduces the autocorrelation inherent to any MCMC chain. If this parameter is not specified at initialization, it is inferred automatically from the sample shape. Separately, to ensure that the chains have reached the stationary distribution $p_{\mu,\theta}$ before any configuration is used to estimate expectation values, the user can also set the number of thermalization sweeps, i.e., the number of initial sweeps that are performed and discarded before any sample is collected. This burn-in step is necessary because the chains are typically initialized from an arbitrary starting configuration, and some number of moves is required before the chain's state becomes representative of the target distribution.

As for the \texttt{ExactSampler}, the class exposes a \texttt{sample} method that returns the sampled configurations, their log-amplitudes, and their associated weights,
\begin{equation}
    \texttt{AbstractMCSampler.sample}: \texttt{num\_samples} \; \longmapsto \; \left(S,\; \log\psi_\theta(S),\; \frac{|\psi_\theta(S)|^{1/\texttt{log\_prob\_factor} - \mu}}{\mathcal{N}}\right),
\end{equation}
where $\mathcal{N} = \sum_{s \in S} |\psi_\theta(s)|^{1/\texttt{log\_prob\_factor} - \mu}$. The parameters \texttt{log\_prob\_factor} and $\mu$ can thus be tuned jointly to implement different importance-sampling schemes. Indeed, since the chain samples from $p_{\mu,\theta}$ rather than the target distribution directly, each configuration is re-weighted by the ratio between the desired probability, $|\psi_\theta(s)|^{1/\texttt{log\_prob\_factor}}$, and the sampling probability, $|\psi_\theta(s)|^{\mu}$, up to normalization. By default, $\mu$ is set to $2$, corresponding to sampling directly from the Born distribution $|\psi_\theta|^2$, and \texttt{log\_prob\_factor} is set to $1/2$. With this choice the importance-sampling exponent $1/\texttt{log\_prob\_factor} - \mu$ vanishes, so every sampled configuration is assigned equal weight, exactly as expected when sampling directly from the target distribution.

Two samplers that inherit from this abstract class are provided: \texttt{MCSampler} and \texttt{MCSamplerCont}. The two differ simply in the type of samples they produce, discrete for the former and continuous for the latter. 

The \texttt{CutoffSampler}, introduced in Ref.~\cite{krinitsinTimedependentVariationalMonte2026}, extends the \texttt{MCSampler} class and alleviates the estimation bias of the conventional approach of sampling from the Born distribution~\cite{sinibaldiUnbiasingTimedependentVariational2023}. It does so by assigning a nonzero probability to every configuration in the Hilbert space, controlled by the cutoff parameter~$\epsilon$.

\subsubsection{Proposer}\label{sec: proposer}
As anticipated, the \texttt{AbstractMCSampler} relies on a proposer, a function that, given a single configuration $s$, proposes a new configuration $s'$ together with the associated proposal ratio. This is implemented by the \texttt{AbstractProposer} interface,
\begin{equation}
    \texttt{AbstractProposer.\_\_call\_\_}: s \longmapsto \left(s',\; \log C(s,s')\right).
\end{equation}
Note that, for numerical stability, the proposer returns the logarithm of the proposal ratio $C(s,s')$ entering the acceptance probability of Eq.~\ref{eq: acceptance ratio}, consistently with the convention adopted for the log-amplitudes discussed above. In addition, a proposer may optionally implement a custom thermalization routine, invoked exclusively during the thermalization sweeps, which allows proposer-specific hyperparameters (such as a step size or a flip probability) to be adaptively tuned, typically by targeting a prescribed acceptance rate.

Our library provides the following proposer classes, ready to be used in combination with \texttt{MCSampler}:
\begin{itemize}
    \item \texttt{SpinFlip}: $s'$ is obtained from $s$ by flipping a single spin, chosen uniformly at random among all sites.
    \item \texttt{SpinFlipZ2}: as in \texttt{SpinFlip}, $s'$ is obtained by flipping a single, randomly chosen spin. In addition, on average once every $30$ proposal steps, all spins are flipped simultaneously, so that global spin-inversion ($\mathbb{Z}_2$) symmetry is enforced in the sampling process.
    \item \texttt{SpinFlipZeroMag}: $s'$ is obtained from $s$ while ensuring that the proposed configuration remains in the zero-magnetization sector, i.e., the sum of all spins in every configuration must vanish. In order to enforce this constraint, a spin-exchange move is proposed rather than a single spin flip. One up spin and one down spin are each selected uniformly at random among the occupied sites and their values are swapped.
\end{itemize}
On the other hand, we provide two different proposers tailored to continuous-variable VMC, to be used in combination with \texttt{MCSamplerCont}:
\begin{itemize}
    \item \texttt{RWM}: implements the random walk Metropolis algorithm \cite{metropolis1953equation}. Given a configuration $s$ representing the positions of all particles in real space, $s'$ is obtained by displacing every particle according to
    \begin{equation}
        s' = s + \delta s, \qquad \delta s \sim \mathcal{N}(0, \Sigma^2),
    \end{equation}
    where $\mathcal{N}(0, \Sigma^2)$ denotes a normal distribution with zero mean and variance $\Sigma^2$, sampled independently for every coordinate of $s$. The step size $\Sigma$ is tuned during thermalization so that the Metropolis--Hastings acceptance rate, given by the average of Eq.~\ref{eq: acceptance ratio} over all produced samples, remains close to a target value.
    \item \texttt{MALA}: implements the Metropolis-adjusted Langevin algorithm \cite{rossky1978brownian}. As in the previous case, the positions of all particles are displaced, but the proposal now also incorporates a drift term along the gradient of the log-amplitude,
    \begin{equation}
        s' = s + \tau\,\nabla\mu\text{Re}\left(\log\psi_\theta(s)\right) + \sqrt{2\tau}\,\xi, \qquad \xi \sim \mathcal{N}(0,1)
    \end{equation}
    where the gradient is taken with respect to the configuration $s$. By biasing proposals towards regions of higher probability density, this drift term typically improves sampling efficiency compared to the undirected random walk of \texttt{RWM}. As in the previous case, the step size $\tau$ can be tuned during thermalization to target a prescribed acceptance rate.
\end{itemize}
Users can readily implement new proposal rules by defining classes that inherit from \texttt{AbstractProposer}, taking care that the call method remains compatible with \texttt{jax.jit}.

\subsubsection{Autoregressive Networks}\label{sec: autoregressive}

Autoregressive networks represent the Born distribution as a product of normalized conditional probabilities~\cite{hibat2020recurrent}.
For an ordered configuration $s=(s_1,\ldots,s_N)$, the distribution and the corresponding pure-state amplitude satisfy
\begin{equation}\label{eq: autoregressive distribution}
    \begin{aligned}
        p_\theta(s)&=\prod_{i=1}^{N}p_\theta(s_i\mid s_{<i}),
        \qquad \sum_{s_i}p_\theta(s_i\mid s_{<i})=1,\\
        \log\psi_\theta(s)&=\tfrac12\log p_\theta(s)+i\varphi_\theta(s),
    \end{aligned}
\end{equation}
where $s_{<i}=(s_1,\ldots,s_{i-1})$ and $\varphi_\theta(s)$ is the phase of the wavefunction.
Drawing the local states successively from these conditionals produces independent configurations at fixed parameters, avoiding MCMC thermalization and autocorrelation between successive samples.

jVMC provides autoregressive models in the form of \texttt{RWKVCPM} in one dimension and \texttt{RWKV2D} in two dimensions.
For \texttt{RWKV2D}, causal connections can follow a rectangular grid or a supplied graph.
The RWKV, or Receptance Weighted Key Value architecture,
retains efficient recurrent sampling and admits parallel evaluation across sites for a known configuration, with a computational cost that grows linearly with sequence length at fixed network width and depth, compared with the quadratic cost of full self-attention~\cite{peng2023rwkv,rigo2025operator}.
As for other networks, a custom autoregressive model must provide a sampling method,
\begin{equation}\label{eq: autoregressive network interface}
    \begin{aligned}
        \texttt{NET.\_\_call\_\_}:\;&s\longmapsto\log\psi_\theta(s),\\
        \texttt{NET.sample}:\;&k\longmapsto s\sim p_\theta,
    \end{aligned}
\end{equation}
where $k$ is a random key and the sampled configuration has the physical shape specified when constructing \texttt{NQS}.
Both methods must be compatible with JAX transformations and use the same parameters and conditional probabilities.
The \texttt{NQS} class handles batching and distribution over devices.

Initialization follows the interface of Sec.~\ref{sec: nqs}: the model is passed to \texttt{NQS} together with the configuration shape, evaluation batch size, and an optional seed for parameter initialization.
The presence of a sampling method identifies the network as a generator, and \texttt{MCSampler} then uses direct sampling without a proposal rule.
For example, the following model generates eight-site configurations using two-site patches:
\begin{lstlisting}[style=pythonstyle]
import jVMC

net_ar = jVMC.nets.RWKVCPM(
    L=8, patch_size=2, hidden_size=8,
    embedding_size=8, num_layers=1, log_prob_factor=0.5
)
psi_ar = jVMC.vqs.NQS(net_ar, 8, batch_size=16, seed=31)
sampler_ar = jVMC.sampler.MCSampler(
    psi_ar, update_proposer=None, key=32, num_samples=64
)
samples_ar, log_psi_ar, weights_ar = sampler_ar.sample()
\end{lstlisting}
The returned arrays contain the configurations, their logarithmic amplitudes, and uniform weights $w_i=1/N_s$ for estimating expectation values from Born-distributed samples.
For pure states, \texttt{log\_prob\_factor} is $0.5$.

A fixed particle number or magnetization can be imposed by masking conditional choices incompatible with the target sector and renormalizing the remaining probabilities~\cite{hibat2020recurrent}.
These masks must be used consistently in amplitude evaluation and sampling.
The implementation provides \texttt{ParticleConserving\allowbreak{}Autoregressive} for \texttt{RWKVCPM}, and the \texttt{fixed\_n\_up} option for binary \texttt{RWKV2D} models; the available constraints depend on the chosen model.

Site patching is likewise implemented inside the network: grouping $b$ sites with local Hilbert-space dimension $d$ gives $d^b$ possible patch configurations at each sampling step, which are decoded into physical site values before being returned.
Symmetry averaging must preserve consistency with direct sampling and is subject to the restrictions described in Sec.~\ref{sec: sym net}.

\subsection{Constructing Observables} \label{sec: observables}

As computing quantum expectation values in VMC reduces to averaging local observables, as described in Sec.~\ref{sec: vmc}, a generic observable object is simply an object that, given a configuration $s$ and the \texttt{NQS}, must be able to produce the corresponding local observable. This is precisely what the general \texttt{AbstractOperator} interface provides,
\begin{equation}
    \texttt{AbstractOperator.get\_O\_loc}: (s,\, \texttt{NQS},\, \xi) \longmapsto O_{\text{loc}}(s, \xi).
\end{equation}
Since an operator may itself depend on additional parameters, such as time, we collectively denote the set of these parameters by $\xi$. From this common abstract interface, two different kinds of operators branch off, as the computation of the local observable is fundamentally different in the two cases: discrete and continuous operators.

\subsubsection{Discrete Operators}

For a generic discrete operator $\hat O$, the local observables can be written as:
\begin{equation}\label{eq: o_loc discrete}
    O_{\text{loc}}(s)=\sum_{s'}\exp\!\left(\log\psi_\theta(s') - \log\psi_\theta(s)\right)\bra{s}\hat O\ket{s'}.
\end{equation}
Notice that, the amplitudes are subtracted in log-space and only exponentiated afterward, to avoid the numerical instabilities that would otherwise arise from evaluating $\psi_\theta(s')/\psi_\theta(s)$ directly when amplitudes vary over many orders of magnitude. Efficiently evaluating Eq.~\ref{eq: o_loc discrete} therefore requires a method to identify, for a given configuration $s$, the set of all configurations $s'$ connected to it, defined as
\begin{equation}
    C_s = \left\{ s' : M(s,s') := \langle s|\hat O|s'\rangle \neq 0 \right\}.
\end{equation}
There exists an important class of operators for which such an algorithm is particularly simple: those that map every configuration $s$ to exactly one connected configuration $s'$ in the same basis, rather than to a set of several connected configurations. These are the so-called branch-free operators. In jVMC they serve as elementary building blocks: products of branch-free operators are again branch-free, and general operators are constructed as sums of such terms. As a concrete example, we construct the operator 
\begin{equation}
    \hat O(t) = \hat\sigma^z_0\left(\hat\sigma^x_0 + 2\hat\sigma^y_1\right) + \cos(t)\,\hat\sigma_0^z\hat\sigma_1^z,
\end{equation}
and inspect its matrix representation, evaluated at a given value of $t$, by calling the \texttt{.to\_dense()} method:

\begin{lstlisting}[style=pythonstyle]
f_t = lambda t, **kwargs: jnp.cos(t)
O = SigmaZ(0) * (SigmaX(0) + 2 * SigmaY(1)) + f_t * SigmaZ(0) * SigmaZ(1)

O.to_dense(basis, t=0.25 * jnp.pi)
\end{lstlisting}

\begin{lstlisting}[style=outputstyle]
matrix([[ 0.70710678+0.j,  0.        -2.j,  1. +0.j, 0.+0.j],
        [ 0.        +2.j, -0.70710678+0.j,  0. +0.j, 1.        +0.j],
        [-1.        +0.j,  0.        +0.j,-0.70710678+0.j, 0.        +2.j],
        [ 0.        +0.j, -1.        +0.j,  0. -2.j, 0.70710678+0.j]])
\end{lstlisting}

This example also illustrates a key design feature of jVMC operators: they support ordinary algebraic composition, sums, products, and multiplication by (possibly parameter-dependent) scalar coefficients, via operator overloading, so that Hamiltonians and other observables can be assembled directly from elementary building blocks, in a syntax that closely mirrors their mathematical expression. Time (or, more generally, $\xi$) dependent coefficients, such as \texttt{f\_t} above, are simply passed as callables that return the coefficient as a function of the relevant parameter.

The connected configurations $C_s$ and the associated matrix elements $M(s,s')$ can be obtained directly by calling \texttt{O.get\_conn\_elements} on a set of configurations $S$ and a given value of $t$. This lower-level method is what \texttt{get\_O\_loc} internally relies on, combining its output with the network amplitudes according to Eq \ref{eq: o_loc discrete}. Importantly, \texttt{get\_conn\_elements} is fully compatible with \texttt{jax.jit}, so that varying $t$, or any other parameter on which the operator depends, does not trigger recompilation. \\

Moreover, discrete operators can also be converted to a sparse matrix representation via the \texttt{.to\_sparse()} method, in direct analogy to \texttt{.to\_dense()} introduced above. This is particularly useful for small to moderately sized systems, where the full Hamiltonian (or any other observable of interest) can still be represented explicitly and diagonalized using standard sparse eigensolvers, e.g., to obtain exact reference energies against which the variational results can be benchmarked.\\

Our library provides a range of predefined elementary operators, covering the most common building blocks used in many-body Hamiltonians:
\begin{itemize}
    \item Spin-$1/2$ operators: \texttt{SigmaX}, \texttt{SigmaY}, \texttt{SigmaZ} (the Pauli matrices), and \texttt{SigmaPlus}, \texttt{SigmaMinus} (the raising and lowering operators $\hat\sigma^\pm = (\hat\sigma^x \pm i\hat\sigma^y)/2$).
    \item Fermionic operators: \texttt{Creation}, \texttt{Annihilation}, and \texttt{Number} ($\hat c^\dagger$, $\hat c$, and $\hat n$, respectively). These can be instantiated as either spinful or spinless. Unlike bare spin operators, they incorporate the appropriate Jordan--Wigner string in their matrix elements, ensuring that the correct fermionic anticommutation relations are respected once the operators are mapped onto the chosen encoding.
\end{itemize}

Beyond these predefined building blocks, users can straightforwardly define custom operators. As an example, we show the explicit definition of $\hat\sigma^y$:

\begin{lstlisting}[style=pythonstyle]
from jVMC.operator.discrete.branch_free import Operator

class SigmaY(Operator):
    def __init__(self, idx):
        super().__init__(
            l_dim=2, idx=idx, diag=False, fermionic=False, spin=None
        )

    @property
    def mat_els(self):
        return np.array([1j, -1j], dtype=global_defs.DT_OPERATORS_CPX)
    
    @property
    def map(self):
        return np.array([1, 0], dtype=jnp.int32)
\end{lstlisting}

In particular, the user must specify: the local Hilbert space dimension on which the operator acts, via \texttt{l\_dim}; whether the operator is diagonal, via \texttt{diag} (enabling further internal optimizations); whether it is fermionic, via \texttt{fermionic}; and, if it is spinful, its spin quantum number, via \texttt{spin}. Since the operator is branch-free, it then suffices to specify, for each of the \texttt{l\_dim} possible values of the local degree of freedom, the single value it is mapped to, through the \texttt{map} property, together with the corresponding matrix element, through the \texttt{mat\_els} property. In the example above, this recovers exactly the action of the Pauli matrix $\hat\sigma^y$ on a spin-$1/2$ degree of freedom: an input value of $0$ is mapped to $1$ with matrix element $\langle 1|\hat\sigma^y|0\rangle = i$, and an input value of $1$ is mapped to $0$ with matrix element $\langle 0|\hat\sigma^y|1\rangle = -i$.

\subsubsection{Continuous Operators}
Continuous operators differ fundamentally from discrete ones, as they do not require finding the set of connected configurations for a given configuration $s$. Indeed, the quantum expectation value of a continuous operator $\hat O$ can be rewritten as
\begin{equation}
    \langle\psi_\theta|\hat O|\psi_\theta\rangle = \int dx\, |\psi_\theta(x)|^2\, \frac{(\hat O\psi_\theta)(x)}{\psi_\theta(x)} := \int dx\, |\psi_\theta(x)|^2\, O_{\text{loc}}(x),
\end{equation}
which has exactly the same structure as the discrete case discussed above: an average of a local observable $O_{\text{loc}}(s)$ over configurations sampled from $|\psi_\theta(s)|^2$, only now $x$ denotes a continuous (e.g., real-space) configuration rather than a discrete one.

Choosing the real-space basis, the local observable of a diagonal operator is simply given by the operator evaluated at the particle positions $x$; no reference to the wavefunction is required at all, since a diagonal operator acts on $|x\rangle$ as a multiplication by a scalar function of $x$ alone. For a non-diagonal operator, by contrast, one must instead specify how the operator acts on the amplitude itself, since $(\hat O\psi_\theta)(x)$ is no longer simply proportional to $\psi_\theta(x)$. As a concrete example, consider the kinetic operator $\hat T$, whose local value is given by
\begin{equation}
	\frac{(\hat T\psi_\theta)(x)}{\psi_\theta(x)} = -\frac{1}{2m}\sum_{i=1}^{P}\left(\nabla_i^2\log\psi_\theta(x) + \left(\nabla_i\log\psi_\theta(x)\right)^2\right),
\end{equation}
where $i$ labels the particle, $P$ is the total number of particles, $m$ is the particle mass, and all gradients and Laplacians are taken with respect to the particles' positions (we set $\hbar = 1$). This expression follows from the identity $\nabla^2\psi_\theta/\psi_\theta = \nabla^2\log\psi_\theta + (\nabla\log\psi_\theta)^2$.

jVMC currently provides the following continuous operators:
\begin{itemize}
    \item Diagonal operators: \texttt{PotentialOperator}, \texttt{CoulombInteraction}, \texttt{ParticleDensity}, and \texttt{Monopole}.
    \item Non-diagonal operators: \texttt{TotalKineticOperator}.
\end{itemize}

As for discrete operators, users can straightforwardly define custom continuous operators. As an example, we show the explicit definition of the potential operator:
\begin{lstlisting}[style=pythonstyle]
from jVMC.operator.continuous.base import Operator

class PotentialOperator(Operator):
    def __init__(self, geometry, potential):
        super().__init__(geometry, is_diagonal=True)
        if not callable(potential):
            raise ValueError('The property potential has to be a function.')
        self._potential = potential
    
    @property
    def potential(self):
        return self._potential
        
    def _get_O_loc(self, s, apply_fun, parameters, kwargs):
        return self.potential(s)
\end{lstlisting}
Here, \texttt{geometry} is a class used by continuous operator to encode the physical setting they act on, and \texttt{potential} is any user-supplied callable implementing the (diagonal) potential energy as a function of the configuration $x$ (for more information see the example in Sec.~\ref{sec: quantum_dot}).

Importantly, the class inherits from the generic \texttt{Operator} base class, so that the user only has to implement \texttt{\_get\_O\_loc} for a single configuration $x$. Internally, the library takes care of vectorizing this computation over an entire batch of configurations, exactly as done for the \texttt{NET} in Sec.~\ref{sec: net}. Note that \texttt{\_get\_O\_loc} receives not the \texttt{NQS} object itself, but rather the network's raw \texttt{apply\_fun} and its \texttt{parameters} directly: this gives non-diagonal operators, such as \texttt{TotalKineticOperator}, direct access to the underlying function being differentiated, which is what is actually needed to construct the automatic-differentiation calls (gradients, Laplacians) entering their local value, without incurring the additional overhead of going through the full \texttt{NQS} wrapper.

\subsubsection{Obtaining Statistics}\label{sec: SampledObs}
Once the local observables have been gathered through the \texttt{get\_O\_loc} method, jVMC provides a simple class to compute statistics over them: \texttt{SampledObs}. This class is initialized with a set of observations and their associated weights, such as those returned by the sampler (see Sec.~\ref{sec: sampler}). Beyond simple statistics such as the mean, variance, and covariance, the class can also compute the autocorrelation time via the fast Fourier transform and the Gelman--Rubin statistic~\cite{vats2021revisiting}, both of which are standard diagnostics for assessing the quality and convergence of an MCMC estimate.

Since the sampler already holds all the ingredients needed to construct these observations, namely the samples and their weights, it can be used directly to obtain the expectation value of an observable $\hat O$, returned as a fully populated \texttt{SampledObs} instance,
\begin{equation}
    \texttt{AbstractSampler.\_\_call\_\_}: \hat O \longmapsto \texttt{SampledObs}\left(O_\text{loc}(S)\right).
\end{equation}
Thus, all the statistics of the resulting estimator are all immediately available as methods of the returned object, without any additional bookkeeping on the user's side. As an example, we compute the magnetization along the $x$ axis for a randomly initialized RBM:
\begin{lstlisting}[style=pythonstyle]
M_x_op = sum(SigmaX(n) for n in range(L)) / L
M_x = sampler(M_x_op)

print('Magnetization:', M_x)
print('Autocorrelation time:', M_x.get_autocorrelation_time(n_chains))
print('Gelman-Rubin statistic:', M_x.get_R_hat(n_chains))
\end{lstlisting}
\begin{lstlisting}[style=outputstyle]
Magnetization: 6.7405e-01+0.0000e+00j $\pm$ 4.0622e-03 (Var = 6.7589e-02)
Autocorrelation time:  8.3453e-01 $\pm$ 2.0593e-01 (Var = 1.6962e-01)
Gelman-Rubin statistic: 1.0003317267261622
\end{lstlisting}

Together with \texttt{SampledObs}, our library provides a closely related object: \texttt{LazySampledObs}. Rather than taking as input a fully materialized set of observations, as \texttt{SampledObs} does, \texttt{LazySampledObs} is built to accept a reusable generator of observations, which it consumes one batch at a time, accumulating the running statistics incrementally as each new batch becomes available. 
This design pairs naturally with \texttt{NQS.lazy\_gradients} (see Sec.~\ref{sec: nqs}): the generator returned by \texttt{lazy\_gradients} can be fed directly into \texttt{LazySampledObs}, so that statistics over the parameter gradients are computed and correctly accumulated batch by batch, without ever materializing the full Jacobian in memory. This is particularly valuable for large networks, where the number of parameters is such that even a single instance of this array would no longer fit in VRAM. As we discuss later, this mechanism is a key ingredient in the efficient implementation of the optimizer (see Sec.~\ref{sec: optimizer}), which relies extensively on gradient statistics of exactly this form.

\subsection{Objective Function}\label{sec: objective function}

Before introducing the algorithms that jVMC provides to optimize the \texttt{NQS} towards a
target state, we define the loss function $\mathcal{L}$ and how its gradient with respect to the
variational parameters $\theta$ is computed. All objective functions implemented in jVMC share a
common structure: $\mathcal{L}$ is the expectation value of a local loss $\ell_\theta(s)$ over the
Born distribution, estimated as a
weighted sum over the configurations $s_i$ and weights $w_i$ held by a sampler,
\begin{equation} \label{eq: local loss}
    \mathcal{L}(\theta) = \mathbb E_{s\sim p_\theta}\left[\ell_\theta(s)\right]\approx \sum_i w_i\,\ell_\theta(s_i).
\end{equation}
The local loss may be complex-valued, but its expectation value is required to be real for all
$\theta$, so that $\mathcal{L}$ is a well-defined loss.

Since complex parameters are always represented as pairs of real parameters
$(\text{Re}\,\theta_k, \text{Im}\,\theta_k)$ (see Sec.~\ref{sec: nqs}), $\theta_k$ denotes a real
parameter throughout. The gradient of Eq.~\eqref{eq: local loss} reads (see Appendix~\ref{app: grad loss fn})
\begin{equation}\label{eq: general gradient}
    \partial_k\mathcal L = 2\,\text{Cov}_p\left(\text{Re}(\Gamma_k),\ell\right) + \expval[p]{\partial_k\ell}.
\end{equation}
The first term stems from the dependence of the sampling distribution on $\theta$, the second from
the explicit dependence of the local loss. As will become clear in Sec.~\ref{sec: optimizer}, it is
convenient to recast Eq.~\eqref{eq: general gradient} as
\begin{equation}\label{eq: grad from force}
    \partial_k\mathcal L = 2\,\text{Re}\left(F_k\right),
\end{equation}
in terms of the complex force vector
\begin{equation}\label{eq: force}
    F_k = \text{Cov}_p\left(\Gamma_k^*,\ell\right) + \frac12 R_k,
    \qquad
    R_k = \expval[p]{\partial_k\text{Re}(\ell)}
    - 2\,\expval[p]{\text{Im}(\Gamma_k)\,\text{Im}(\ell)} \in \mathbb R.
\end{equation}
The covariance is the standard VMC force, while $R_k$ collects all remaining contributions; it
vanishes whenever $\ell$ is the local estimator of a $\theta$-independent Hermitian operator. \\

The common interface for all loss functions is the \texttt{AbstractObjectiveFunction}, which
exposes two methods: \texttt{\_\_call\_\_}, which evaluates the local losses $\ell_\theta(s_i)$ on
the configurations currently held by a sampler and returns them as a \texttt{SampledObs}, and
\texttt{value\_and\_grad}, which additionally computes the Jacobian and the force vector,
\begin{equation}
    \texttt{AbstractObjectiveFunction.value\_and\_grad}: \texttt{AbstractSampler} \longmapsto \left(\ell_\theta(S),\; \Gamma,\; F\right).
\end{equation}
The computation of $F$ can be skipped via the boolean flag \texttt{compute\_grad}. At
construction, the flag \texttt{batched\_jacobian} determines whether $\Gamma$ is returned as a
\texttt{SampledObs} or as a \texttt{LazySampledObs} (see Sec.~\ref{sec: SampledObs}); the latter is
preferable whenever the full Jacobian does not fit in memory.

The objective functions provided by jVMC differ only in the choice of $\ell_\theta$, which
determines the form of $R_k$:
\begin{itemize}
    \item \texttt{Observable}: $\mathcal{L}=\langle\hat O\rangle$ is the quantum expectation value
    of a Hermitian operator $\hat O$ (Sec.~\ref{sec: observables}), and the local loss is the local
    observable $O_\text{loc}(s)$ introduced in
    Sec.~\ref{sec: vmc}. In this case $R_k=0$ (see Appendix~\ref{app: grad loss fn}), and
    the force vector takes the standard form
    \begin{equation}\label{eq: force observable}
        F_k = \text{Cov}_p\left(\Gamma_k^*,\ O_\text{loc}\right).
    \end{equation}

    \item \texttt{Estimator}: the local loss is an arbitrary user-defined, real-valued function of
    the configuration and the parameters,
    \begin{equation}
        \ell_\theta(s) = f(s,\theta) \in \mathbb{R},
    \end{equation}
    without any operator structure. Since $\text{Im}(\ell)=0$, Eq.~\eqref{eq: force} gives
    $R_k=\expval[p]{\partial_{\theta_k} f}$, so both terms of Eq.~\eqref{eq: general gradient}
    contribute and
    \begin{equation}
        F_k = \text{Cov}_p\left(\Gamma_k^*,\ f\right) + \frac12\expval[p]{\partial_{\theta_k} f},
    \end{equation}
    where $\partial_{\theta_k} f$ is obtained via automatic differentiation. Regularization terms
    such as $f = \lambda\|\theta\|^2$ are a special case: since $f$ does not depend on $s$, the
    covariance vanishes identically and $F_k=\lambda\theta_k$.

    \item \texttt{ParametricObservable}: $\mathcal{L}=\langle\hat O(\theta)\rangle$ is the quantum
    expectation value of an operator $\hat O(\theta)$ that is Hermitian for all $\theta$ and whose
    matrix elements depend explicitly on the variational parameters, in addition to the implicit
    dependence through $\psi_\theta$:
    \begin{equation}
        \ell_\theta(s) = O_\text{loc}(s;\theta) = \sum_{s'} O_{ss'}(\theta)\,\frac{\psi_\theta(s')}{\psi_\theta(s)} .
    \end{equation}
    The implicit dependence is treated exactly as for \texttt{Observable}, while the explicit one
    yields $R_k=\expval[p]{(\partial_{\theta_k}\hat O)_\text{loc}}$ (see
    Appendix~\ref{app: grad loss fn}), so that
    \begin{equation}
        F_k = \text{Cov}_p\left(\Gamma_k^*,\ O_\text{loc}\right)
        + \frac12\expval[p]{(\partial_{\theta_k}\hat O)_\text{loc}},
        \qquad
        (\partial_{\theta_k}\hat O)_\text{loc}(s) = \sum_{s'}\partial_{\theta_k}O_{ss'}(\theta)\,\frac{\psi_\theta(s')}{\psi_\theta(s)} .
    \end{equation}
    Since $\partial_{\theta_k}\hat O$ is itself Hermitian, its expectation value is real. The
    derivatives $\partial_{\theta_k}O_{ss'}(\theta)$ are obtained by automatic differentiation of a
    user-supplied function encoding the explicit parameter dependence of the matrix elements.
    \item \texttt{Infidelity}: $\mathcal{L}=\mathcal{I}$ is the infidelity between the trial state
    $|\psi_\theta\rangle$, sampled by the primary sampler, and a reference state $|\chi\rangle$,
    sampled independently by a second, dedicated \texttt{reference\_sampler}. Optionally, a unitary
    operator $\hat O$ can be applied to the reference state, so that
    \begin{equation}
        \mathcal{I} = 1 - \frac{\langle\psi_\theta|\hat O|\chi\rangle\langle\chi|\hat O^\dagger|\psi_\theta\rangle}{\langle\psi_\theta|\psi_\theta\rangle\,\langle\chi|\chi\rangle}.
    \end{equation}
    This construction is particularly useful when $|\chi\rangle$ is a previously optimized state and $\hat O$ an evolution operator, in which case minimizing $\mathcal{I}$ implements projected time-dependent VMC (ptVMC)~\cite{gravina2025neural}. The infidelity is estimated by combining two local ratios, each sampled from only one of the two distributions,
    \begin{equation}
        \mathcal{I}^\psi_\text{loc}(s) = \frac{(\hat O\chi)(s)}{\psi_\theta(s)},\; s\sim|\psi_\theta|^2,
        \qquad
        \mathcal{I}^\chi_\text{loc}(s) = \frac{(\hat O^\dagger\psi_\theta)(s)}{\chi(s)},\; s\sim|\chi|^2,
    \end{equation}
    whose expectation values are $\langle\psi_\theta|\hat O|\chi\rangle/\langle\psi_\theta|\psi_\theta\rangle$
    and $\langle\chi|\hat O^\dagger|\psi_\theta\rangle/\langle\chi|\chi\rangle$, respectively. The
    local loss is
    \begin{equation}
        \ell_\theta(s) = 1 - \mathcal{I}^\psi_\text{loc}(s)\,\langle \mathcal{I}^\chi_\text{loc}\rangle_\chi.
    \end{equation}
    This is precisely the local estimator of the $\theta$-independent Hermitian operator
    \begin{equation}
        \hat H_\chi= 1-\hat O|\chi\rangle\langle\chi|\hat O^\dagger/\langle\chi|\chi\rangle,
    \end{equation}
    
    i.e.\ $\mathcal{I}=\langle\hat H_\chi\rangle$. Therefore, as for \texttt{Observable}, $R_k=0$ and the force vector reduces to a single covariance,
    \begin{equation}
        F_k = \text{Cov}_p\left(\Gamma_k^*,\ \ell\right)
        = -\,\text{Cov}_p\left(\Gamma_k^*,\ \mathcal{I}^\psi_\text{loc}\right)\langle \mathcal{I}^\chi_\text{loc}\rangle_\chi .
    \end{equation}

    It has been shown that the signal-to-noise ratio of the infidelity estimator vanishes as $\mathcal{I}\to 0$, which makes them unreliable indicators of
    convergence~\cite{gravina2025neural}. To mitigate this, \texttt{Infidelity} supports two control-variate variance-reduction schemes, selected via the
    \texttt{control\_variate} flag~\cite{gravina2025neural}. In one of the examples provided in jVMC, we show how to use this objective function to construct a simple ptVMC optimization loop. Note that the same objective can be used for
    non-unitary $\hat O$. In that case, the properly normalized infidelity is obtained by
    multiplying the fidelity term above by
    \begin{equation}
        c = \frac{\langle\chi|\chi\rangle}{\langle\chi|\hat O^\dagger\hat O|\chi\rangle},
    \end{equation}
    which does not depend on the variational parameters $\theta$. The two objectives therefore
    share the same minimizer, and their gradients differ only by the constant factor $c$. All
    the estimators above remain valid, since $\hat O|\chi\rangle\langle\chi|\hat O^\dagger$ is
    Hermitian for any $\hat O$. 
\end{itemize}

As in the previous cases, users are free to define custom objective functions by implementing classes that inherit from \texttt{AbstractObjectiveFunction}, simply providing a \texttt{\_\_call\_\_} and a \texttt{value\_and\_grad} method with the signatures defined above. This architectural choice gives the user full freedom to go beyond the objective functions provided out of the box, for example by adding regularization terms to an existing loss. Crucially, since \texttt{AbstractObjectiveFunction} sits at the same level of abstraction as \texttt{Observable}, \texttt{Estimator}, and \texttt{ParametricObservable}, any such custom objective function integrates seamlessly with the rest of the library, in particular with the optimizers discussed next (see Sec.~\ref{sec: optimizer}), without requiring any changes downstream.

\subsection{The Stepper} \label{sec: stepper}
As it will become clear in Sec.~\ref{sec: optimizer}, different ground-state search and (real- or imaginary-)time dynamics algorithms require solving a first-order differential equation for the set of variational parameters, of the form
\begin{equation}
    \frac{d\theta}{dt} = f(t,\theta),
\end{equation}
where the specific functional form of $f$ is determined by the chosen algorithm. Our library therefore provides a collection of different time integrators, which we refer to as steppers. These are divided into two groups:
\begin{itemize}
    \item \textbf{Simple steppers}: \texttt{Euler} and \texttt{Heun}, which advance the solution using a fixed, user-specified time step $dt$.
    \item \textbf{Adaptive steppers}: \texttt{AdaptiveHeun}, \texttt{RK23}, and \texttt{RK45}, which instead adjust the time step $dt$ dynamically over the course of the integration.
\end{itemize}

Adaptive steppers work by computing, at every integration step, two candidate solutions of different accuracy, $\theta_1$ and $\theta_2$, and measuring the distance between them under a chosen metric $\mathcal{M}$. If this distance exceeds a prescribed tolerance, meaning
\begin{equation} \label{eq: adaptive condition}
    \|\theta_1-\theta_2\|_\mathcal{M} > \texttt{atol} + \texttt{rtol}\,\|\theta\|_\mathcal{M},
\end{equation}
the time step is rejected and reduced, and two new candidate solutions are recomputed with the smaller step. This process repeats until either the tolerance is satisfied or, if provided, a minimum allowed time step is reached. The three adaptive steppers differ in how the two candidate solutions $\theta_1,\theta_2$ are obtained. For \texttt{AdaptiveHeun}, the higher-accuracy solution is obtained by performing two consecutive half-sized Heun steps in place of a single full step, so that estimating the error requires solving the underlying differential equation twice, at a correspondingly higher computational cost. \texttt{RK23} and \texttt{RK45}, by contrast, are embedded Runge--Kutta pairs, the Bogacki--Shampine and Dormand--Prince schemes respectively, in which a single set of intermediate stage evaluations of $f$ is reused to construct two solutions of different order (second and third order for \texttt{RK23}; fourth and fifth order for \texttt{RK45}) via two different linear combinations of the same stages. The higher-order solution of the pair is then the one actually propagated forward.

While users can implement custom norms for $\mathcal{M}$, our library already provides two: a simple root-mean-square norm, and a norm that uses the quantum geometric tensor $S$ as its metric, so that the distance between the two solutions is evaluated directly in the Hilbert space (see Sec.\ref{par: gs search}).

\subsection{The Optimizer} \label{sec: optimizer}
The optimizer class is responsible for taking all the key modules defined so far and orchestrating them into a complete, runnable training algorithm. Our library defines the general structure of an optimizer through the \texttt{AbstractOptimizer} class. Given the current set of parameters $\theta$, this class implements a full VMC loop, producing an updated set of parameters $\theta'$ as output. This is done through three steps: (1) the sampler is used to draw a new set of configurations $S$; (2) these configurations are passed to the chosen objective function via its \texttt{value\_and\_grad} method (see Sec.~\ref{sec: objective function}); and (3) the resulting output is passed to \texttt{AbstractOptimizer.get\_update()}, which returns $\theta'$. While this three-step structure is entirely general, the specific rule used to turn the objective function's output into a parameter update is algorithm-dependent, and must therefore be implemented separately by each concrete optimizer, by overriding \texttt{get\_update}.

This three-step procedure is precisely what computes $f(t,\theta)$ in the differential equation of Sec.~\ref{sec: stepper}, and a single stepper call may require evaluating $f$ several times within one integration step. Each such evaluation would require drawing an entirely new set of configurations from the distribution $p_\theta$ associated with that intermediate $\theta$, at significant additional computational cost. To avoid this, \texttt{AbstractOptimizer} provides the option to draw new configurations only once, at the very first stage of a stepper call, and to reuse this same set of configurations for all subsequent intermediate stages, correcting for the fact that they were drawn from a different distribution than the current one via importance sampling, as proposed in Ref.~\cite{wiersema2026numerical}. This simple technique reduces the computational cost of otherwise expensive multi-stage time integrators, and it moreover reduces the statistical noise entering the error estimate of Eq.~\ref{eq: adaptive condition}, allowing for a more reliable dynamical refinement of the time step. Whether this scheme is used can be decided at initialization through the flag \texttt{resample\_stepper}.

This general interface also implements two convenience utilities, \texttt{ground\_state\_search} and \texttt{time\_evolution}, both of which simply repeat the VMC loop described above: for a fixed number of iterations $N$ in the former case, or until a target time $t_\text{max}$ is reached in the latter. Both methods can additionally take, as optional arguments, a list of observables and a list of callbacks; when provided, the corresponding quantum expectation values are computed and stored at every iteration, and the given callbacks are invoked, allowing the user to monitor, log, or otherwise intervene in the optimization as it progresses.

As with the other core components of the library, this class is designed so that users can easily implement custom optimizers by inheriting from \texttt{AbstractOptimizer} and implementing only \texttt{get\_update}, making it straightforward to develop and experiment with new optimization algorithms. Our library already implements several optimizers tailored to ground-state search and to dynamics, which we discuss in what follows. Importantly, all the optimizers provided in jVMC are compatible with the $\texttt{LazySampledObs}$ interface (see Sec \ref{sec: SampledObs}), so that if required the algorithms never materialize the full Jacobian to construct quantities such as the quantum geometric tensor or the tangent kernel.

\subsubsection{Ground State Search Optimizers} \label{par: gs search}
The ground state of a Hamiltonian can be approximated using the variational principle, which guarantees that the quantum expectation value of the Hamiltonian over any trial wavefunction $|\psi_\theta\rangle$ is always greater than or equal to the true ground-state energy $E_0$,
\begin{equation}\label{eq: var principle}
    E_\theta = \frac{\langle\psi_\theta|\hat H|\psi_\theta\rangle}{\langle\psi_\theta|\psi_\theta\rangle} \geq E_0.
\end{equation}
As a consequence, minimizing the left-hand side of Eq.~\ref{eq: var principle} with respect to the parameters $\theta$ provides a way to approximate the ground state of the system. Since this optimization problem is in general non-convex, several strategies can be adopted in practice to reliably converge to a good approximation of the ground state.

Our library implements a simple \texttt{Adam} optimizer, which drives the optimization towards the ground state through an adaptive variant of stochastic gradient descent (SGD)~\cite{kingma2014adam}, using the gradient $\nabla_\theta E_\theta$ obtained from the \texttt{Observable} objective function introduced in Sec.~\ref{sec: objective function}. \\

Since plain SGD ignores both the geometry of the variational manifold and the gauge redundancy
typically present in overparametrized networks (multiple values of $\theta$ can represent the
same physical state $|\psi_\theta\rangle$), jVMC also provides stochastic reconfiguration (SR) as
an optimization scheme~\cite{sorella1998green, becca2017quantum}. SR can be understood as a
natural-gradient method~\cite{amari1998natural}. Rather than measuring the step size in parameter
space, it measures it through the Fubini--Study distance between the normalized states
$|\psi_\theta\rangle$ and $|\psi_{\theta+\delta\theta}\rangle$, which to second order reads
$\sum_{kl}\delta\theta_k\,\text{Re}[S_{kl}]\,\delta\theta_l$ with the quantum geometric tensor (QGT) defined as
\begin{equation}\label{eq: qgt}
    S_{kl}=\text{Cov}_p\left(\Gamma_k^*,\Gamma_l\right).
\end{equation}
Minimizing the linearized loss at fixed step length yields the update
$\theta\to\theta+\eta\,\delta\theta$, with $\delta\theta$ the solution of
\begin{equation}\label{eq: sr equation}
    \text{Re}[S]\,\delta\theta = -\text{Re}[F],
\end{equation}
where $F$ is the force vector of Eq.~\eqref{eq: force} and the factor $\tfrac12$ relating
$\text{Re}[F]$ to $\nabla_\theta\mathcal L$ has been absorbed into the learning rate $\eta$. As gauge redundancies manifest themselves as (near-)zero eigenvalues of $\text{Re}[S]$, before solving Eq.~\eqref{eq: sr equation} this has to be regularized. A standard approach is to shift and rescale the diagonal of the QGT
\begin{equation}
    S \to S + \lambda_1 I + \lambda_2\,\text{diag}(S),
\end{equation}
where $\text{diag}(S)$ denotes the matrix that agrees with $S$ on the diagonal and vanishes elsewhere. The two regularization strengths can be provided to the \texttt{SR} class at initialization, as \texttt{diagonal\_shift} and \texttt{diagonal\_scale} respectively, either as fixed scalars or as functions of the current optimization step, allowing for e.g. an annealed regularization schedule.

Since the QGT has shape $N_p \times N_p$, solving Eq.~\ref{eq: sr equation} directly can become computationally prohibitive as the network size increases. In this regime, the \texttt{MinSR} optimizer can be used instead: it targets the same update as SR, but exploits the push-through identity to recast the solution of Eq.~\ref{eq: sr equation} as the inversion of an $N_s \times N_s$ matrix, the neural tangent kernel \cite{rende2024simple, chen2024empowering}. As for the QGT, the tangent kernel can be regularized in exactly the same manner, again via the \texttt{diagonal\_shift} and \texttt{diagonal\_scale} parameters. Both SR and MinSR can be combined with any objective function provided in jVMC, but they are geometrically motivated only when $\mathcal L$ depends on $\theta$ solely through $|\psi_\theta\rangle$.
 
\subsubsection{Real Time Evolution}
The time-dependent Schr\"odinger equation can be approximately solved within VMC by means of time-dependent VMC (tVMC): the parameters of the network are promoted to time-dependent quantities $\theta(t)$, chosen at every instant so that the family of variational states $|\psi_{\theta(t)}\rangle$ follows the exact solution as closely as possible. While $|\psi_{\theta(t)}\rangle$ is constrained by construction to the variational manifold defined by the network, the exact time-evolved state will in general leave this manifold at every infinitesimal time step. All tVMC schemes therefore rest on the same idea, the time-dependent variational principle (TDVP): the Schr\"odinger equation is projected onto the tangent space of the manifold at the current point $\theta(t)$~\cite{hackl2020geometry}.

The tangent space is spanned by the vectors 
\begin{equation}
    |V_k\rangle = \big(\partial_{\theta_k} - \langle\partial_{\theta_k}\rangle\big)|\psi_\theta\rangle
\end{equation}
so that $\langle s|V_k\rangle = \big(\Gamma_k(s) - \langle\Gamma_k\rangle\big)\psi_\theta(s)$, where the mean is subtracted to remove changes of norm and global phase. As mentioned above restricting the time derivative to the manifold, leaves a residual 
\begin{equation}
    |R\rangle = \sum_k \dot\theta_k\,|V_k\rangle + i\big(\hat H - E_\theta\big)|\psi_\theta\rangle .
\end{equation}
The TDVP requires that $|R\rangle$ has no component along the vectors $\ket{V_k}$ spanning the tangent space: 
\begin{equation}\label{eq: tdvp projection}
    \langle V_k|R\rangle = \sum_{k'} S_{kk'}\,\dot\theta_{k'} + i F_k=0,
\end{equation}
with the quantum geometric tensor $S$ and the force vector $F$, the same Monte Carlo estimators that enter Eq.~\ref{eq: sr equation}, now keeping both real and imaginary parts.

If the network is holomorphic, the tangent space is a complex vector space and the full complex condition $\langle V_k|R\rangle = 0$ can then be imposed, leading to the Dirac--Frenkel form
\begin{equation} \label{eq: tdvp dirac frenkel}
    S\,\dot\theta = -iF .
\end{equation}

If the network is not holomorphic, the tangent space is only a real vector space, therefore not closed under multiplication by $i$. This leads to an additional degree of freedom: the test vectors can be chosen as either $|V_k\rangle$ or $i|V_k\rangle$, leading to different implementations of the same variational principle.
\begin{itemize}
    \item \textbf{McLachlan form}: the test vectors are chosen as $\ket{V_k}$, so that
    \begin{equation}\label{eq: tdvp mclachlan}
        \text{Re}[S]\,\dot\theta = \text{Re}[-iF].
    \end{equation}
    It can be shown that this formulation is not symplectic.

    \item \textbf{Lagrangian form}: the test vectors are chosen as $i\ket{V_k}$, so that
    \begin{equation}\label{eq: tdvp lagrangian}
        \text{Im}[S]\,\dot\theta = \text{Im}[-iF].
    \end{equation}
\end{itemize}

Our library implements all three variants of the same variational principle within a single \texttt{TDVP} class. When the network is holomorphic, the class automatically uses the Dirac--Frenkel form. When the network is not holomorphic, the user selects between the McLachlan and the Lagrangian form via the boolean flag \texttt{make\_real}, which determines whether the real or the imaginary part of $S$ and $-iF$ is used to construct the linear system solved at each time step. \\

tVMC generally requires a more delicate numerical treatment than ground state search. In ground-state search only the final, converged state matters: transient inaccuracies in $S$ or $F$ along the optimization path slow down convergence but do not bias the state eventually reached, in much the same way that noisy gradients still converge on average in ordinary stochastic gradient descent. In dynamics, by contrast, every intermediate step directly determines the trajectory being traced out, and errors in $S$ and $F$ at a given step propagate forward and accumulate over the subsequent evolution, exactly as for the right-hand side of any numerically integrated ODE. This leads to two practical considerations. First, $S$ and $F$ must be resolved with considerably higher statistical accuracy than is typically required for ground-state search, which in practice means using a substantially larger number of samples $N_S$ at every step. 

Second, the naive diagonal regularization of Sec.~\ref{sec: optimizer}, harmless for ground-state search since it only reshapes the descent path towards the same fixed point, can introduce a substantial bias in the dynamics. Shifting and rescaling $S$ changes the metric defining the tangent-space projection itself, and therefore changes which state on the manifold is actually being tracked at every step, rather than merely how quickly it is reached. Consequently, more elaborate regularization schemes, tailored specifically to the linear systems arising in tVMC, are required; we discuss the solvers implemented in our library for this purpose in Sec.~\ref{sec: solvers}. 

\subsubsection{Solvers}\label{sec: solvers}

As explained in Sec.~\ref{sec: optimizer}, most optimizers require the solution of a linear system of the form $A x = b$ at every iteration. Our library abstracts this step behind a common \texttt{AbstractSolver} interface
\begin{equation}
 \texttt{AbstractSolver.\_\_call\_\_}: (A, b) \longmapsto (x, \texttt{info}),
\end{equation}
where \texttt{info} is a dictionary of solver-specific diagnostics (e.g. the residual of the solve, or the condition number of $A$) that can be logged at every step via the callback mechanism of \texttt{AbstractOptimizer} (Sec.~\ref{sec: optimizer}). Since $A$ is always either the QGT or the tangent kernel, both Hermitian positive semi-definite matrices by construction (Sec.~\ref{sec: optimizer}), our library provides the following solvers, all of which exploit this structure:

\begin{itemize}
    \item \textbf{Pinv}: computes a truncated Moore--Penrose pseudoinverse of $A$ via its eigendecomposition $A = V\,\text{diag}(\lambda)\,V^\dagger$ (exploiting Hermiticity, rather than a general SVD), and solves the system as $x = V\,\text{diag}(\lambda^{-1})\,V^\dagger b$, without ever explicitly assembling the pseudoinverse matrix itself. Eigenvalues satisfying $|\lambda_i/\lambda_\text{max}| \leq \texttt{pinv\_cutoff}$ are treated as zero and dropped, using a hard cutoff fixed at initialization.

    \item \textbf{PinvSNR}: extends \texttt{Pinv} with two complementary, smooth (rather than hard) regularization filters, applied to the inverse eigenvalues, following the general pseudoinverse regularization scheme of \cite{medvidovic2023variational, schmitt2020quantum}. The first replaces the fixed hard cutoff of \texttt{Pinv} with a smooth cutoff function of $|\lambda_i/\lambda_\text{max}|$. This avoids the discontinuous jumps in the resulting update over the course of the optimization, a concern that matters most in dynamics. Moreover, rather than being fixed, the cutoff itself is searched adaptively at every single linear solve: starting from a loose value, it is decreased until either the residual of the solve falls below a target tolerance \texttt{pinv\_tol}, or a minimum allowed cutoff \texttt{pinv\_cutoff} is reached, whichever comes first. Note that this differs from \texttt{Pinv}'s use of the same parameter name: for \texttt{Pinv}, \texttt{pinv\_cutoff} is the (fixed) cutoff itself, whereas for \texttt{PinvSNR} it is only the floor of an adaptively determined cutoff.

    The second filter is activated by setting \texttt{snr\_tol} to a nonzero value at initialization. It further suppresses update directions along eigenvectors of $A$ whose associated component of the force vector $b$ cannot be statistically resolved from sampling noise, based on the signal-to-noise ratio (SNR) of $b$ projected onto the eigenbasis of $A$~\cite{schmitt2020quantum}. Computing this SNR requires the densely materialized, per-sample log-derivative Jacobian (i.e. a \texttt{SampledObs} rather than a \texttt{LazySampledObs}, obtained by setting \texttt{batched\_jacobian=False} on the objective function, see Sec.~\ref{sec: objective function}) together with the per-sample local estimator entering $b$. For the same reason, SNR regularization is only meaningful for \texttt{SR} and \texttt{TDVP}, whose linear system is formulated directly in the $p$-dimensional parameter space spanned by the per-sample Jacobian; \texttt{MinSR}, which instead operates in the $N_s$-dimensional sample-space eigenbasis of the tangent kernel, must therefore be used with \texttt{snr\_tol}~$=0$.

    \item \textbf{CG}: an iterative conjugate-gradient solver, wrapping \texttt{jax.scipy.sparse.linalg.cg}. Unlike \texttt{Pinv} and \texttt{PinvSNR}, which require $A$ to be diagonalized and therefore explicitly materialized as a dense matrix, \texttt{CG} only ever needs $A$ through matrix-vector products, and can therefore also be used with a purely functional, matrix-free representation of $A$.
\end{itemize}

All solvers that require diagonalizing $A$ additionally support three diagonalization backends. These can be selected via $\texttt{diagonalization\_mode}$: $\texttt{"device"}$, using $\texttt{jax.numpy.linalg.eigh}$ on a single device; $\texttt{"host"}$, using $\texttt{numpy.linalg.eigh}$ on the CPU; and $\texttt{"distributed"}$, which diagonalizes $A$ jointly across all available devices using JAXMg~\cite{wiersema2026jaxmg}. The distributed mode allows matrices considerably larger than what a single device could otherwise handle to be diagonalized, without ever materializing the full matrix on every device simultaneously, thereby avoiding out-of-memory errors that would otherwise limit the size of the QGT, or of the tangent kernel, that can be inverted. We demonstrate in the benchmarking section below, that the parallelization moreover yields speedup compared to single device runtime for large problem sizes.

\subsection{Using Multiple GPUs}\label{sec: multiple gpus}

As outlined throughout the previous sections, jVMC is designed from the ground up to run efficiently across multiple GPUs, possibly spread across different nodes. Concretely, this means that: samples are generated using independent Markov chains distributed in parallel across all available devices (Sec.~\ref{sec: sampler}); every network evaluation and gradient computation, whether triggered directly by the \texttt{NQS} or internally by an operator's local-value computation (Sec.~\ref{sec: nqs}, Sec.~\ref{sec: observables}), is likewise carried out in parallel across devices; the diagonalization of the QGT or the tangent kernel, required by \texttt{SR}, \texttt{TDVP}, and \texttt{MinSR}, is parallelized across devices via JAXMg (Sec.~\ref{sec: solvers}); and, even when a finite \texttt{batch\_size} restricts how many configurations are processed at once, each individual batch is itself sharded so that it lives across multiple devices rather than being placed on a single one. Correspondingly, the QGT and the tangent kernel are both built and stored distributed across devices, so that no single device ever holds the full matrix (see Appendix~\ref{app: distributed Jacobian products}). 

This is achieved by building on the sharding mechanism natively provided by JAX. A one-dimensional device mesh, with a single axis named \texttt{"devices"}, is constructed once at import time from all globally visible devices. Throughout the library, arrays of shape $(N_s,...)$ consistently follow the convention of being distributed along this leading axis, so that the samples living on a given device are also the ones processed on that same device at every subsequent stage of the computation, without requiring data movement between devices in between. For this scheme to divide evenly, quantities such as the number of samples or the number of Markov chains must be divisible by the number of devices. The library handles this automatically by increasing the number of samples or Markov chains if necessary.

For the specific case of matrix diagonalization via JAXMg (Sec.~\ref{sec: solvers}), a second, two-dimensional device mesh, with axes labeled \texttt{"row"} and \texttt{"col"}, is constructed separately from the one-dimensional mesh used elsewhere, with devices arranged in as close to a square grid as possible. This reflects the fact that efficient distributed diagonalization algorithms typically require a genuinely two-dimensional partitioning of the matrix being diagonalized, rather than the one-dimensional, sample-indexed sharding used for the rest of the pipeline.\\

Beyond sharding across the devices of a single process, jVMC also supports running across multiple processes, for instance to scale a computation across several nodes of a cluster, each contributing its own set of GPUs. By default, the library assumes a single-process setting. Multi-process execution is enabled by setting the global flag \texttt{global\_defs.JVMC\_USE\_DISTRIBUTED} to \texttt{True} before importing the library. In this case, jVMC calls \texttt{jax.distributed.initialize()} at start-up, which establishes the communication between processes and merges their devices into a single, globally addressable mesh spanning all nodes. On most common cluster environments, such as SLURM, JAX detects the process layout automatically, so this default call is sufficient and no further configuration is needed. On systems where this automatic detection is not available, the user must instead call \texttt{jax.distributed.initialize()} themselves, with a custom configuration, before importing jVMC. The flag \texttt{global\_defs.JVMC\_USE\_DISTRIBUTED} must still be set to \texttt{True} in this case, so that the library operates in multi-process mode. On a SLURM environment, when using multiple nodes, one process per device should be used. JAXMg requires one process per device also on a single node.

\subsubsection{Sharded Decorator}
The parallelization of per-configuration computations is centralized in a single reusable building block, the \texttt{sharded} decorator. It turns a plain method acting on one batch of configurations into a function that is vectorized and distributed across the device mesh. The decorator first vectorizes the method over the configurations, and then distributes it in one of two ways. With \texttt{automatic\_sharding=True}, the vectorized method is simply compiled with \texttt{jax.jit}, and JAX's sharding propagation places the computation according to the sharding of the inputs. Otherwise, it is distributed explicitly with \texttt{jax.shard\_map}, using user-specified \texttt{in\_specs} and \texttt{out\_specs} that fix how each argument and the output are partitioned across the mesh. The first option requires no additional information, while the second gives precise control over the partitioning. In addition, the decorator handles batching transparently whenever a \texttt{batch\_size} smaller than the number of samples is requested: the configurations are padded to the next multiple of \texttt{batch\_size}, split into batches, and each batch is sharded across the devices before being passed to the distributed function, without any communication between devices. When constructed with \texttt{yield\_iter=True}, rather than eagerly evaluating and concatenating all batches, the decorator instead returns a lazily evaluated, reusable generator over per-batch results. This is precisely the mechanism underlying \texttt{NQS.lazy\_gradients} and its use throughout \texttt{LazySampledObs} (Sec.~\ref{sec: nqs}, Sec.~\ref{sec: SampledObs}), allowing gradients (or any other per-configuration quantity) to be both distributed across devices and streamed batch by batch, without ever materializing the full result on any single device at once. Users can further control how this decorator is constructed by setting the global flag \texttt{global\_defs.JVMC\_BATCH\_LOOP} to either \texttt{"python"} or \texttt{"compiled"} before importing the library. The flag determines how the loop over batches is executed, and the two options trade runtime performance against recompilation cost. In the \texttt{"python"} mode, the loop is a plain Python \texttt{for} loop that repeatedly calls a single jit-compiled function. In the \texttt{"compiled"} mode, the loop itself is expressed in JAX and compiled together with the batched computation. Compiling the loop generally yields better performance, since the compiler can fuse operations across iterations. However, it requires recompilation every time the shape of the input changes, i.e., whenever the number of samples changes. Besides the compilation time, this has a second, less obvious cost: JAX caches every compiled executable, and for large networks these executables are themselves big. Repeatedly changing the number of samples therefore accumulates a growing set of cached programs, which can by itself exhaust memory and cause out-of-memory errors, even when a single evaluation would comfortably fit. The Python loop avoids both problems: the inner compiled function is always called on inputs of fixed shape \texttt{(batch\_size, \ldots)}, regardless of the total number of configurations, so recompilation, and the accompanying growth of the cache, is only triggered when \texttt{batch\_size} itself is changed. The \texttt{"compiled"} mode is therefore preferable when the number of samples is fixed throughout a run, while the \texttt{"python"} mode is better when this number varies between calls. When imported, the library defaults to \texttt{"python"}. \\

Because this machinery is exposed as a simple decorator, adding a new sharded, batched method to a custom \texttt{NQS}, \texttt{Operator}, or objective function (Sec.~\ref{sec: nqs}, Sec.~\ref{sec: observables}, Sec.~\ref{sec: objective function}) requires no explicit knowledge of \texttt{shard\_map} or manual batching logic on the user's part.
\section{Benchmarks} \label{sec: benchmarks}
\subsection{Parallelism on multiple GPUs}

\begin{figure}[t]
    \centering
    \includegraphics[width=\linewidth]{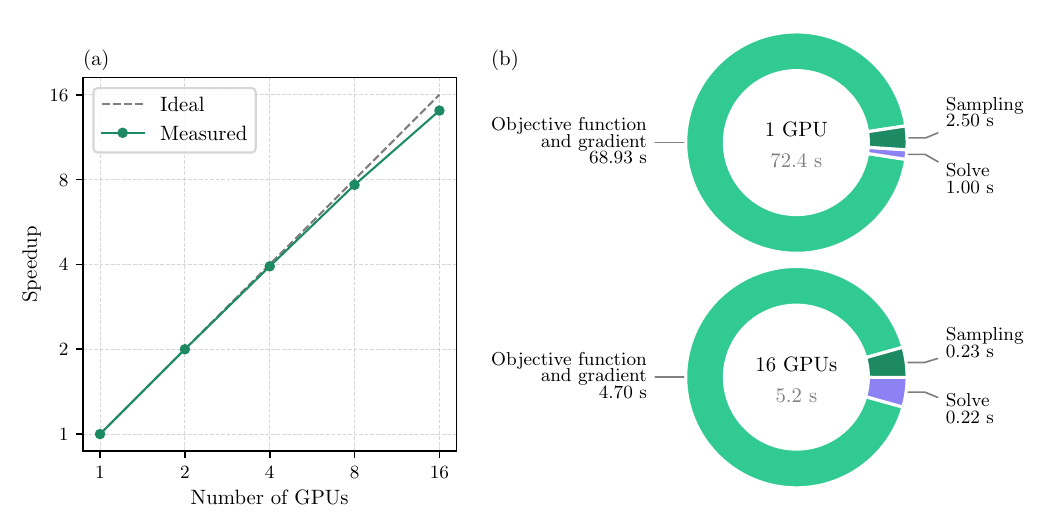}
    \caption{\label{fig: gpu speedup} (a) Speedup of a TDVP step as a function of the number of GPUs used, and (b) breakdown of the time spent in the three main parts of the algorithm. The system is a transverse-field Ising chain with $N=100$ spins, and the variational state is a RNN with $N_p = 2082$ parameters. The number of samples is $N_s = 2^{19}$ and the batch size is $N_b = 2^{10}$. The GPUs are NVIDIA A100 80GB.}
\end{figure}

We benchmark the parallelization capabilities of the library across accelerators with a single step of the TDVP algorithm. Fig.~\ref{fig: gpu speedup}(a) shows an example of an observed speedup when increasing the number of GPUs from 1 to 16. Each node contains only 4 GPUs, but the parallelization is seamless across nodes thanks to the JAX sharding mechanism. Fig.~\ref{fig: gpu speedup}(b) shows the breakdown of the time spent in three parts of the algorithm: the sampling, the computation of the energy and its gradient, and solving the linear system. The first two parts can be fully parallelized across GPUs, while the last part contains the inversion of the quantum geometric tensor, which cannot be sped up in the same way. Note that the inversion can be parallelized with JAXMg, as explained in Sec.~\ref{sec: multiple gpus}, and can provide a speedup in certain cases. Its main goal, however, is to reduce the memory footprint of the inversion. This example, however, does not use JAXMg, as the number of parameters is small enough that the inversion of the QGT is not a memory bottleneck. Nevertheless, the solve step in Fig.~\ref{fig: gpu speedup}(b) still becomes faster with more GPUs, from $1.00$\,s to $0.22$\,s, because it also includes the construction of the QGT from the Jacobian, which is distributed across devices like the rest of the computation.

In this example, most of the time is spent on computing the energy and its gradient. This is due to the large system size: each sample has $100$ connected elements, making the local energy computation expensive. This computation, however, is embarrassingly parallel, and therefore well suited for GPU parallelization.

\subsection{Batching}\label{sec: batching vs memory}

\begin{figure}[t]
    \centering
    \includegraphics[width=\linewidth]{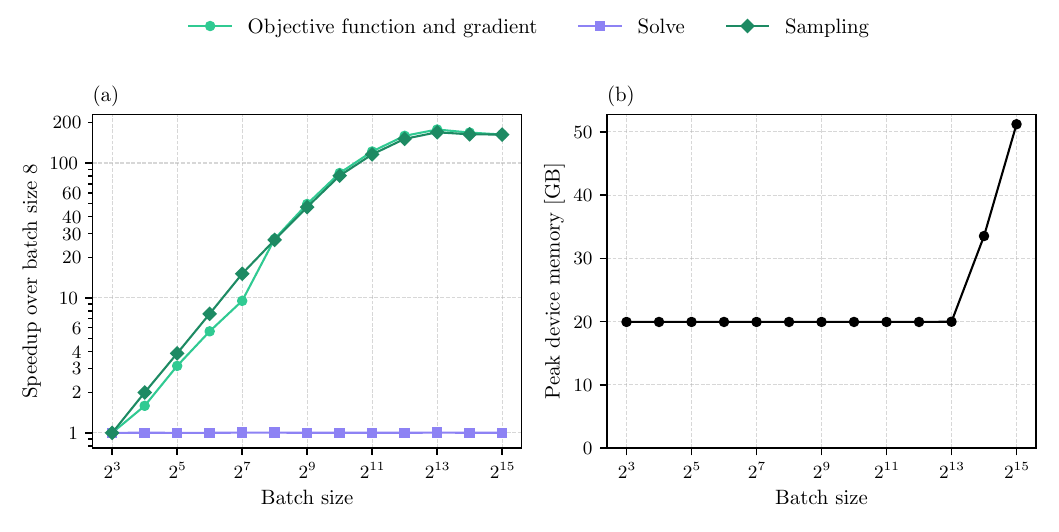}
    \caption{\label{fig: batch size speedup} (a) Speedup as the batch size is increased for the three main parts of the algorithm, and (b) the corresponding memory usage for a single TDVP step. The system is a transverse-field Ising chain with $N=50$ spins, and the variational state is a RWKV with $N_p = 32768$ parameters. The number of samples is $N_s = 2^{12}$ and the GPUs are NVIDIA A100 80GB.}
\end{figure}

The batch size must be carefully chosen to balance the memory usage and the speed of the computation. Fig.~\ref{fig: batch size speedup}(a) shows an example of the speedup for a single TDVP step as a function of the batch size, broken down into the three main parts of the algorithm: sampling, computing the energy and its gradient, and solving the linear system. The first two parts benefit the most from increasing the batch size, with a speedup spanning two orders of magnitude. Notice, however, that the speedup plateaus as the batch size approaches the total number of samples used, while the memory usage in Fig.~\ref{fig: batch size speedup}(b) increases sharply. There is therefore a sweet spot, dependent on the problem, where memory usage can be lowered without sacrificing performance. The cost of solving the linear system, on the other hand, is often dominated by the inversion of the QGT, which is independent of the batch size, so no speedup is observed. 

\subsection{Parallelized Monte Carlo sampling}

\begin{figure}[t]
    \centering
    \includegraphics[width=\linewidth]{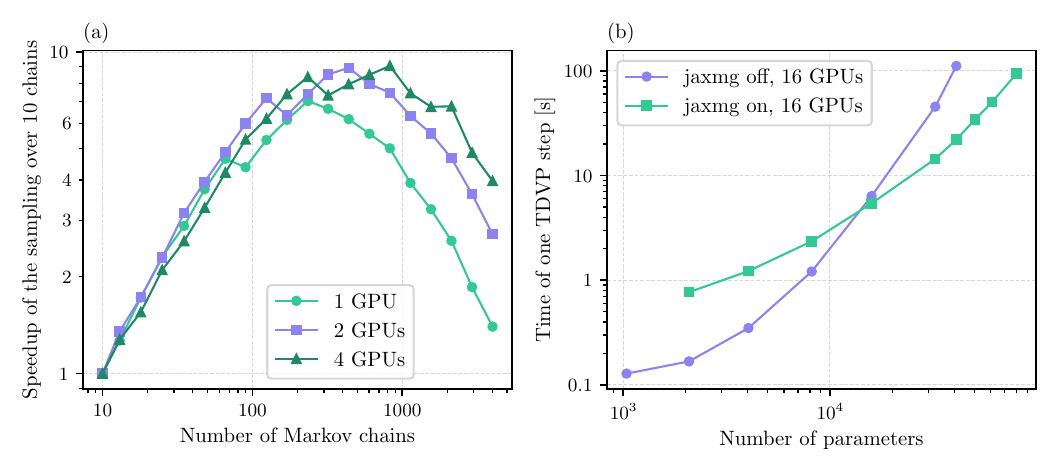}
    \caption{\label{fig: mc chains speedup} (a) Speedup in the sampling time as the number of Monte Carlo chains is increased. The system is a transverse-field Ising chain with $N=50$ spins, and the variational state is a complex CNN with $N_p = 3268$ parameters. The number of samples is $N_s = 2^{12}$, and the GPUs are NVIDIA A100 80GB. The sampling consists of single-spin-flip Metropolis updates with sweeps of $50$ proposals and $20$ thermalization sweeps. The speedup is computed with respect to 10 chains for each curve. (b) Comparison of the execution time of a single TDVP step with and without JAXMg. The curves stop when the simulation runs out of memory. The system is a transverse-field Ising chain with $N=50$ spins, and the variational state is a RNN. The number of samples is $N_s = 2^{12}$ and the batch size is $N_b = 2^{10}$. The GPUs are NVIDIA A100 80GB.}
\end{figure}

Monte Carlo sampling can be parallelized by running multiple Markov chains in parallel, with the chains themselves parallelized across GPUs. Fig.~\ref{fig: mc chains speedup}(a) shows an example of the speedup in sampling time as a function of the number of Markov chains used, on one or more GPUs. At first, increasing the number of chains leads, as expected, to a speedup. At some point, however, the speedup saturates, and then reverses. This happens as the GPU's computing resources become saturated by the parallel network evaluations. Furthermore, each chain has a fixed number of thermalization sweeps, meaning that adding more chains increases the total number of sweeps, leading to a slowdown. Adding more GPUs provides more computational resources and allows more chains to run in parallel before reaching saturation. The optimal number of chains is therefore problem dependent, and should be chosen carefully to maximize the speedup.

\subsection{Distributed matrix inversion with JAXMg}

JAXMg allows the inversion of the QGT to be distributed across multiple GPUs, allowing the use of larger networks for TDVP, or more samples for MinSR. As an example, Fig.~\ref{fig: mc chains speedup}(b) shows the total time of a single TDVP step on 16 GPUs with and without JAXMg. Without JAXMg, the inversion is performed on one GPU, and the maximum number of parameters reached before running out of memory is $N_p = 40862$, while with JAXMg, the maximum number of parameters reached is $N_p = 79886$.

Fig.~\ref{fig: mc chains speedup}(b) also shows that the distributed inversion is slower than the single GPU inversion for small matrices, but becomes faster for larger matrices as the memory usage approaches saturation. JAXMg decomposes the matrix into tiles whose size can be controlled with the tile size parameter $T_A$. By default, we fix this parameter at $T_A = 256$. More information about the tiling can be found in Ref.\cite{wiersema2026jaxmg}.

\subsection{Comparison with NetKet}
\begin{figure}[t!]
    \centering
    \includegraphics[width=\linewidth]{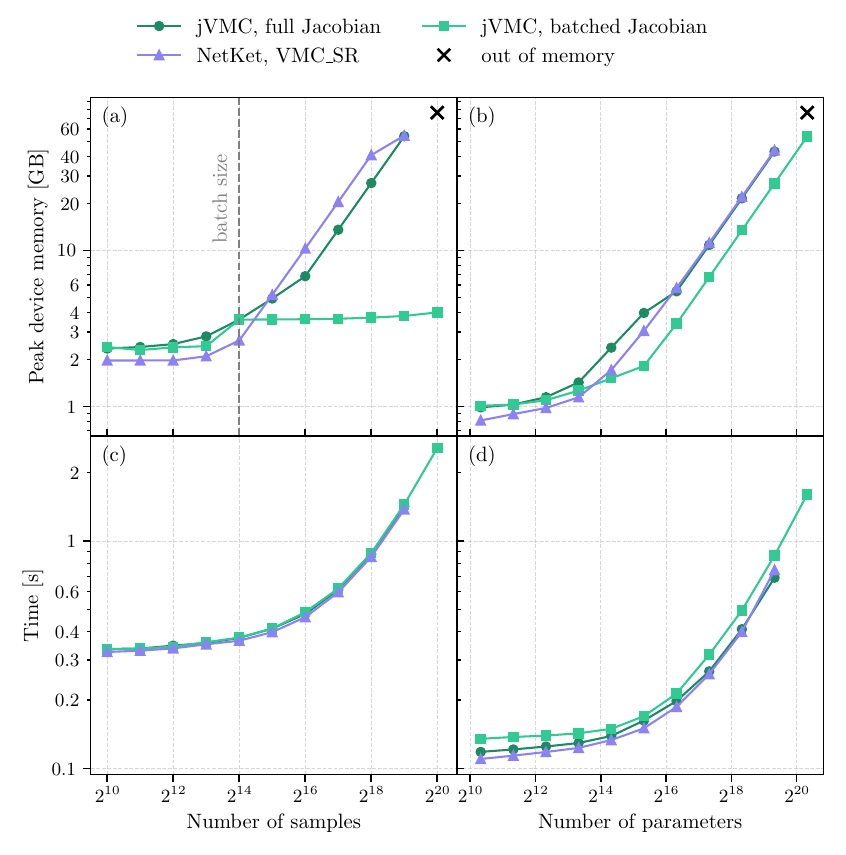}
    \caption{Peak device memory and time of one optimization step in jVMC and NetKet on a single NVIDIA GH200 GPU. (a, c) SR at a fixed number of parameters, $N_p = 6400$, as a function of the number of samples $N_s$. (b, d) MinSR at a fixed number of samples, $N_s = 4096$, as a function of $N_p$. The variational state is a RBM for the transverse-field Ising chain with $N = 80$ spins, in double precision, with $80$ hidden units in (a, c) and $16$ to $16384$ hidden units in (b, d). Both libraries use the same parameters, the same uniformly distributed samples and the same linear solver: an eigendecomposition-based pseudo-inverse with diagonal shift $10^{-3}$ and relative eigenvalue cutoff $10^{-14}$. NetKet runs its \texttt{VMC\_SR} driver, with the neural tangent kernel for MinSR. jVMC builds the Jacobian either in full or in batches of $2^{14}$ samples in (a, c) (dashed line) and $2^{10}$ samples in (b, d); the other series process all samples at once. A step comprises the local energies, the Jacobian, the matrix $S$ (SR) or the neural tangent kernel (MinSR) and the solution of the linear system; sampling is excluded so that the same set of samples can be used in both libraries. Times are obtained after one compilation and one warm-up step. The two libraries give the same parameter update to within a relative difference of $5\times10^{-12}$.}
    \label{fig: netket sr minsr comparison}
\end{figure}

NetKet \cite{vicentini2021netket} is another widely used open-source Python package implementing VMC with NQS, and it is therefore natural to benchmark the two codebases against each other. Before doing so, we note two points. First, both libraries are built on JAX and rely on just-in-time compiled functions, so we expect comparable performance and memory usage for sampling, expectation values and gradients, as confirmed by the benchmarks below. Second, both libraries support multi-GPU and multi-node execution through JAX sharding and handle the distribution of data internally, so that little knowledge of the device layout is required from the user.

One may therefore ask why two libraries with overlapping functionality should be maintained. We see two reasons. First, independent implementations of the same methods provide a direct test of reproducibility, which is particularly valuable in a field where results and computational performance can depend sensitively on implementation details. Second, and more importantly, the two packages follow different design philosophies. NetKet aims at broad coverage of models, Hilbert spaces, operators and algorithms within a general and extensible framework. jVMC instead deliberately keeps its core minimal. It implements a small set of essential building blocks and keeps the path from a high-level call to the underlying JAX operations short and explicit. Since VMC with NQS is still a rapidly evolving field, with few established guidelines on which methods work reliably, we consider this an advantage for method development. New algorithms can be implemented with little code, without navigating extensive abstraction layers, and on the same footing as the built-in functionality. \\

We benchmark the two libraries in three settings, using a restricted Boltzmann machine (RBM) as the variational state.
\begin{enumerate}
    \item We compare the execution time for sampling and for estimating the expectation value of an observable. As observable we take the Hamiltonian of the one-dimensional transverse-field Ising model with $N=80$ spins,
    \begin{equation}\label{eq: Ising Hamiltonian}
        \hat H = -\sum_{i=1}^{N-1}\hat\sigma^z_i\hat\sigma^z_{i+1}-\sum_{i=1}^{N}\hat\sigma^x_i .
    \end{equation}
    The results are shown in Tab.~\ref{tab: netket comparison}, and as expected the two libraries share very similar execution time for both tasks.
    \begin{table}[t]
        \centering
        \begin{tabular}{lcc}
            \toprule
            & Sampling [ms] & $\langle \hat H \rangle$ [ms] \\
            \midrule
            jVMC                              & 49.4 & 5.19 \\
            NetKet, \texttt{IsingJax}         & 46.6 & 4.07 \\
            NetKet, \texttt{LocalOperatorJax} & 46.6 & 7.53 \\
            \bottomrule
        \end{tabular}
        \caption{Execution time of the two stages of a Monte Carlo estimate of $\langle \hat H \rangle$ in jVMC and NetKet on a single NVIDIA GH200 GPU, for the transverse-field Ising chain of Eq.~\eqref{eq: Ising Hamiltonian} with $N = 80$ spins. The variational state is a RBM with $80$ hidden units in double precision; both libraries use identical network parameters. \emph{Sampling}: $4096$ samples from $1024$ Markov chains, single-spin-flip Metropolis updates with sweeps of $80$ proposals and $20$ thermalization sweeps. \emph{Estimate of $\langle \hat H \rangle$}: the full estimate on these samples, including matrix elements, network evaluations, mean and statistical error, with a chunk size of $4096$ network evaluations. The two NetKet rows differ only in the operator implementation and share the same sampler.}
        \label{tab: netket comparison}
    \end{table}

    \item We measure the peak memory usage and the execution time of a single SR step as a function of the number of samples, again for the Hamiltonian in Eq.~\eqref{eq: Ising Hamiltonian}. In both libraries the linear system is solved with the Moore--Penrose pseudoinverse. For NetKet we use \texttt{QGTJacobianDense}. For jVMC we run the benchmark twice, once with the full Jacobian and once with the batched Jacobian. The results are shown in Fig.~\ref{fig: netket sr minsr comparison} (a,c). With the full Jacobian, the two libraries show very similar memory usage and execution time. With the batched Jacobian, the peak memory saturates as the number of samples grows, at the cost of a modest increase in execution time. This makes the batched Jacobian essential for avoiding out-of-memory errors at large sample sizes. To our knowledge, NetKet offers no equivalent option that bounds the memory of the Jacobian while still producing the dense matrix required by direct linear solvers. Its matrix-free alternative avoids storing the Jacobian, but it is designed for iterative solvers.
    \item We repeat the experiment for MinSR, now varying the number of variational parameters at a fixed number of samples. The results are shown in Fig.~\ref{fig: netket sr minsr comparison} (b,d). Again, the two libraries perform very similarly. Since the batched Jacobian is split along the sample axis, each batch still contains all parameters, so its memory grows with the number of parameters and no plateau appears in this case. Nevertheless, the batched Jacobian requires less memory than the full one and runs successfully at parameter counts where both the full-Jacobian implementation of jVMC and NetKet run out of memory.
\end{enumerate}

As a final note, unlike jVMC, the current version of NetKet (3.22.4) does not keep the quantum geometric tensor sharded across devices: every device computes the full $N_p\times N_p$ contribution of its own samples, and these are all-reduced into a complete copy of $S$ on each device. The same holds for the tangent kernel of MinSR: NetKet redistributes the Jacobian along the parameter axis, which splits the computation of $T$ across devices, but $T$ is again all-reduced into a full $N_s\times N_s$ copy on every device. A distributed solver such as JAXMg therefore cannot reduce the memory per device in NetKet, whereas jVMC builds both matrices block by block with reduce-scatter operations, so that neither is ever stored in full on a single device (see Appendix~\ref{app: distributed Jacobian products}).
\section{Examples}\label{sec: examples}

\subsection{The $J_1$--$J_2$ Model Using Mixed Precision}
\label{sec:mixed_precision}

\begin{figure}[t!]
    \centering
    \includegraphics[width=\linewidth]{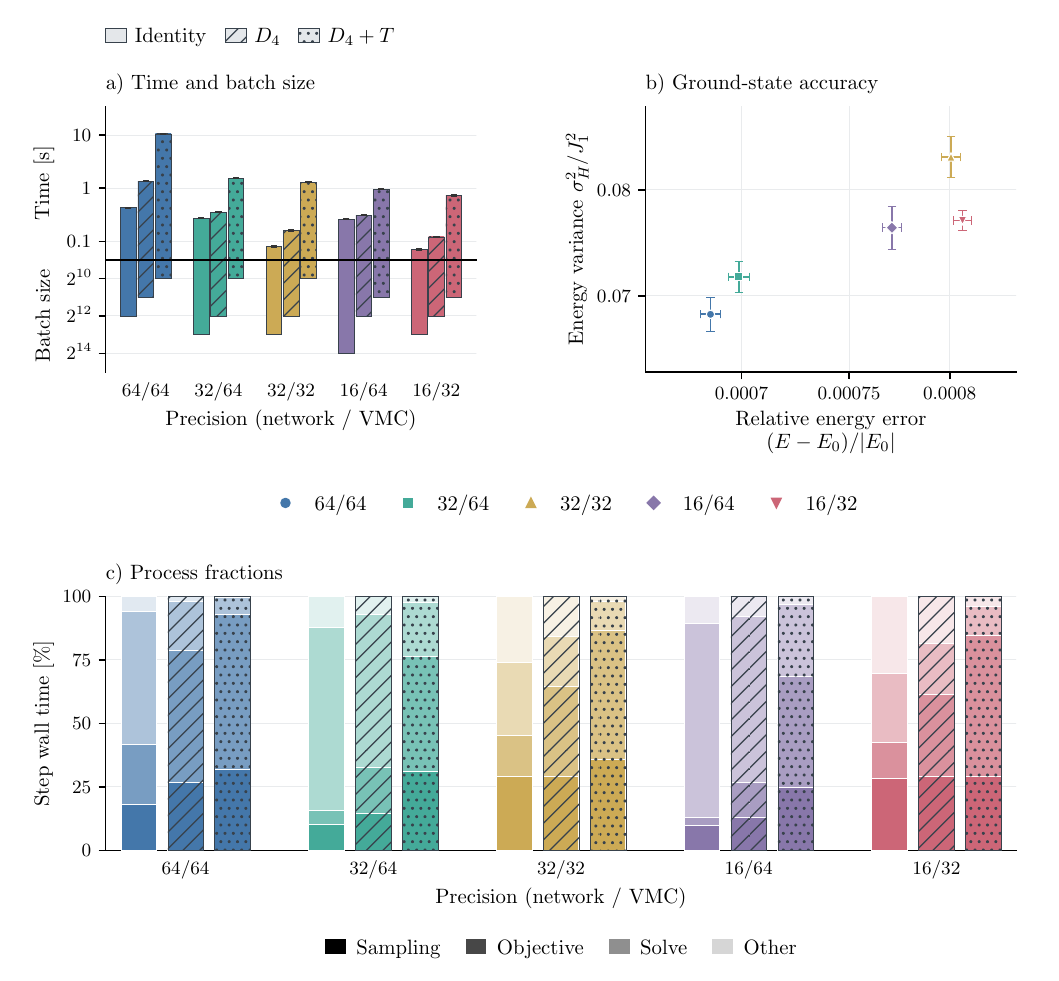}
    \caption{
    Mixed-precision ground-state search for the periodic $6\times6$ $J_1$--$J_2$ model at $J_2/J_1=0.5$, with $N_p=10\,032$ ViT parameters.
    Precision labels give the network and VMC bit widths, respectively, and $T$ denotes lattice translations.
    \textbf{(a)} Step times and evaluation batch sizes for three choices of symmetry.
    Each time is the median over three updates following one warm-up update and includes sampling, objective evaluation, and the solve.
    Whiskers span the observed minimum and maximum, and the lower bars give batch sizes on a separate logarithmic scale.
    \textbf{(b)} Relative energy error and energy variance after $10\,000$ updates with $D_4+T$ symmetry.
    We use the rounded exact-diagonalization energy $E_0=-18.13716\,J_1$ and estimate $\sigma_H^2$ from the sample variance of the complex local energy.
    Horizontal error bars give one Monte Carlo standard error of the energy divided by $|E_0|$.
    Vertical error bars give one standard error of the variance from $4\,000$ whole-chain bootstrap resamples stratified by ensemble.
    \textbf{(c)} Fractions of step wall time spent on sampling, objective evaluation, the solve, and other work, obtained from mean process times over the same three updates as in (a).
    The solve includes construction of the tangent kernel and right-hand side, the linear solve, and projection of the solution back into parameter space.
    }
    \label{fig:mixed_precision_benchmark}
\end{figure}

As a first example, we use the mixed-precision interface of Sec.~\ref{sec: nqs} to approximate the ground state of the spin-$1/2$ $J_1$--$J_2$ Heisenberg model on the square lattice,
\begin{equation}
    \hat H = J_1 \sum_{\langle i,j\rangle} \hat{\mathbf S}_i\cdot\hat{\mathbf S}_j
           + J_2 \sum_{\langle\langle i,j\rangle\rangle} \hat{\mathbf S}_i\cdot\hat{\mathbf S}_j ,
    \label{eq: j1j2}
\end{equation}
where $\langle i,j\rangle$ and $\langle\langle i,j\rangle\rangle$ denote nearest- and next-nearest-neighbour pairs, respectively. At $J_2/J_1 = 0.5$ the two antiferromagnetic couplings compete strongly, the model is highly frustrated, which makes it one of the standard benchmarks for NQS. We consider a $6\times6$ lattice with periodic boundary conditions, for which the reference energy can be found using exact diagonalization. We restrict the search to the sector of zero total magnetization, $S^z_\mathrm{tot}=0$, which contains the ground state. This constraint is enforced during sampling by the \texttt{SpinFlipZeroMag} proposer (Sec.~\ref{sec: proposer}).

Our ansatz follows the vision transformer (ViT) of Viteritti et al.~\cite{viteritti2025square}, whose attention weights depend on the relative displacement of the patches.
We divide the lattice into four $3\times3$ patches, embed each patch in a vector of dimension $12$, and use six transformer layers with four attention heads each, giving $10\,032$ real parameters.
In the modified readout we apply layer normalization to the sum of the patch embeddings and separately normalize the real and imaginary inputs to the complex $\log\cosh$ function.
We work in the Marshall basis, in which the nearest-neighbor spin-exchange matrix elements acquire a minus sign.
The ViT is fully compatible with any abelian symmetry, which here is the square point group $D_4$ and translation.
We impose only the latter on the network through symmetry projector of Sec.~\ref{sec: sym net}.
 
The cost of a VMC step is dominated by network evaluations: the forward passes during sampling, the amplitudes of all configurations connected by $\hat H$ that enter the local energies, and the Jacobian $\Gamma$. These operations tolerate reduced precision well. By contrast, the accumulation of Monte Carlo statistics might be sensitive to round-off errors. jVMC sets the two precisions independently, through arguments of \texttt{NQS}: \texttt{sampler\_dtype}, \texttt{eval\_dtype} and \texttt{grad\_dtype}.
These fix the arithmetic inside the network, while the global output data types fix the precision of everything downstream (Sec.~\ref{sec: nqs}). For instance, a run with a complex network in FP32 and the remaining VMC operations in FP32 is set up as
\begin{lstlisting}[style=pythonstyle]
import os
os.environ["JVMC_DT_OUT_REAL"] = "f32"
os.environ["JVMC_DT_OUT_CPX"] = "c64"

import jax.numpy as jnp
import jVMC
from jVMC.operator.discrete import SigmaZ, SigmaPlus, SigmaMinus

L = 6
net = jVMC.nets.ViterittiViT(
    Lx=L, Ly=L, patch_size=3, layers=6, embed_dim=12, heads=4,
    phase_scale=1.0, param_dtype=jnp.float32, compute_dtype=jnp.float32,
)
psi = jVMC.vqs.NQS(
    net, 
    (L, L), 
    batch_size=16, 
    seed=1234,
    sampler_dtype=jnp.float32,
    eval_dtype=jnp.float32,
    grad_dtype=jnp.float32
)
chains = 8
checkerboard = jnp.indices((L, L)).sum(axis=0) % 2
sampler = jVMC.sampler.MCSampler(
    psi, jVMC.propose.SpinFlipZeroMag(), key=4321,
    init_state=jnp.tile(checkerboard, (chains, 1, 1)),
    num_chains=chains, num_samples=32, sweep_steps=L*L,
    thermalization_sweeps=2,
)

# Periodic Heisenberg bonds in the Marshall basis
H = 0
bonds = [(1, 0, 1.0), (0, 1, 1.0), (1, 1, 0.5), (1, -1, 0.5)]
for y in range(L):
    for x in range(L):
        i = y * L + x
        for dx, dy, J in bonds:
            j = ((y + dy) % L) * L + (x + dx) % L
            sign = -1 if (dx + dy) % 2 else 1
            H += (J / 4) * SigmaZ(i) * SigmaZ(j)
            H += (sign * J / 2) * SigmaPlus(i) * SigmaMinus(j)
            H += (sign * J / 2) * SigmaMinus(i) * SigmaPlus(j)

energy = jVMC.objective_function.Observable(H)
opt = jVMC.optimizer.MinSR(sampler, psi, diagonal_shift=1e-3)
stepper = jVMC.stepper.Euler(0.01)
history = opt.ground_state_search(3, energy, stepper)
print(float(history["energy"]["mean"][-1]))
\end{lstlisting}

For this example we compare each network precision (FP64, FP32, or FP16) with FP64 VMC arithmetic and also test FP32 and FP16 networks with FP32 VMC arithmetic.
Each update uses $2\,000$ samples drawn from $2\,000$ Markov chains with $144$ proposals per sweep.
We optimize the parameters with SPRING~\cite{goldshlager2024kaczmarz} at zero momentum ($\mu=0$), using a diagonal shift $\lambda=10^{-3}$ and unit weight for the projector onto the constant mode of the tangent kernel.
We use a constant learning rate $\eta=0.05$ and bound the Euclidean norm of each parameter step by $0.5$.

We assess runtime and accuracy for the various precision combinations on a single NVIDIA GeForce RTX 5090. First we evaluate the runtime for three symmetry settings: no projection, projection onto the $D_4$ point group, and $D_4$ combined with lattice translations. Lowering the precision also reduces the memory required per configuration, so for every precision pair and symmetry setting we select the evaluation batch size independently, while the gradient batch always contains all $2\,000$ samples. The timings in Fig.~\ref{fig:mixed_precision_benchmark}(a) therefore combine the cheaper arithmetic with the larger batches it allows, which reflects how mixed precision is used in practice. Reduced precision lowers the step time for every symmetry setting. For example, with $D_4$ and translations, FP16/FP32 is $14.4$ times faster than FP64/FP64. This large factor partly reflects the limited double-precision throughput of consumer GPUs. Fig.~\ref{fig:mixed_precision_benchmark}(c) shows how the step time is distributed among sampling, objective evaluation and the linear solve.

To assess how reduced precision affects the final variational state, we train each precision pair for $10\,000$ updates on a single NVIDIA A100 80\,GB, with both $D_4$ and lattice-translation symmetry enforced. The cumulative training time decreases from $27.8$\,h for FP64/FP64 to $15.5$\,h for FP16/FP32. Figure~\ref{fig:mixed_precision_benchmark}(b) shows the energy and variance of each final state, measured independently of training from $10^7$ fresh configurations. For each state, we initialize two independent ensembles of $1\,000$ chains, discard $512$ thermalization sweeps, and retain $5\,000$ configurations per chain. We evaluate the local energies at the configured precision and accumulate the estimator statistics in double precision. Across the five precision pairs, the relative energy error ranges from $0.686\times10^{-3}$ for FP64/FP64 to $0.807\times10^{-3}$ for FP16/FP32.

\subsection{Continuous Space: Electrons in a Quantum Dot}
\label{sec: quantum_dot}
\begin{figure}[t]
    \centering
    \includegraphics[width=\linewidth]{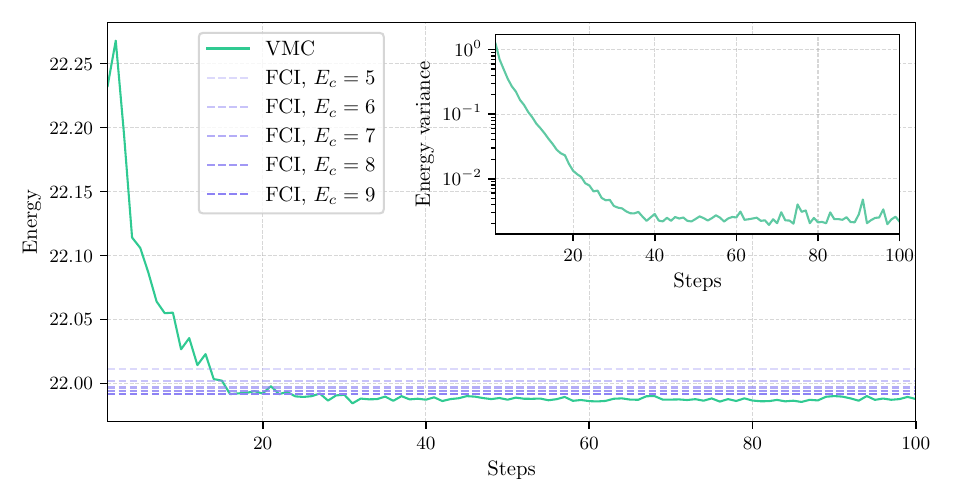}
    \caption{Ground-state search for $N=6$ electrons in the quantum dot of Eq.~\ref{eq: quantum dot hamiltonian} with $\omega = 1$. Main panel: variational energy $\langle H \rangle$ as a function of the optimization step for the ansatz in Eq.~\ref{eq: quantum dot ansatz}, compared with the FCI reference with increasing number of basis functions (dashed lines). The orbital basis includes all Fock--Darwin states up to the cutoff energy $E_c = 6\omega$; the Jastrow factor and the backflow transformation use $E_c = 4\omega$. In total, the ansatz has $827$ complex variational parameters. Inset: energy variance $\sigma^2 = \langle H^2 \rangle - \langle H \rangle^2$, which vanishes for an exact eigenstate, as a function of the optimization step.}
    \label{fig: quantum dot}
\end{figure}

The same variational workflow applies beyond lattice models.
We illustrate it with the ground-state search for $N$ interacting spin-polarized electrons confined in a two-dimensional harmonic trap, a standard model of semiconductor quantum dots~\cite{anisimovas1998energy}. In effective atomic units, $\hbar = m^{*} = e^{2}/4\pi\varepsilon = 1$, the Hamiltonian reads
\begin{equation}\label{eq: quantum dot hamiltonian}
    H = \sum_{i=1}^{N} \left( -\frac{1}{2} \nabla_i^2
        + \frac{\omega^2}{2} r_i^2 \right)
      + \sum_{i<j} \frac{1}{|\mathbf{r}_i - \mathbf{r}_j|},
\end{equation}
where $\mathbf{r}_i \in \mathbb{R}^2$ is the position of the $i$-th electron, $r_i = |\mathbf{r}_i|$, and $\omega$ sets the confinement strength. We consider $N=6$ electrons at $\omega = 1$. This system is small enough that full configuration interaction (FCI) gives a reference energy, and its strong correlation still makes it a nontrivial test.

The Hamiltonian is assembled from predefined operators:

\begin{lstlisting}[style=pythonstyle]
N = 6
n_dim = 2
omega = 1.0

geometry = jVMC.geometry.HyperRectangle(N, n_dim, PBC=False, extent=(6., 6.))
hamiltonian = (
    op.TotalKineticOperator(geometry)
    + op.PotentialOperator(geometry, lambda x: 0.5 * omega ** 2 * jnp.sum(x ** 2))
    + op.CoulombInteraction(geometry)
)
\end{lstlisting}

We here introduce the \texttt{geometry} object which
is mainly needed to define the boundary condition of the problem, and therefore how to compute the distance between particles, and how to initialize the Markov chains. 
A natural single-particle basis for this problem is given by the eigenstates of the noninteracting Hamiltonian, the Fock--Darwin orbitals
\begin{equation} \label{eq: fock darwin}
    \varphi_{\alpha}(r, \theta)
    = \sqrt{\frac{\omega\, n!}{\pi\,(n+|m|)!}}\,
      \bigl(\sqrt{\omega}\, r\bigr)^{|m|}\,
      L_n^{|m|}\!\left(\omega r^2\right)\,
      e^{-\omega r^2/2}\, e^{i m \theta},
    \qquad \alpha \equiv (n, m),
\end{equation}
with radial quantum number $n \in \mathbb{N}$, angular momentum
$m \in \mathbb{Z}$, generalized Laguerre polynomials $L_n^{|m|}$, and single-particle energies $\epsilon_\alpha = \omega\,(2n + |m| + 1)$.

Our ansatz is a variant of the one proposed in Ref.~\cite{nys2024ab}. It combines a single Slater determinant of backflow orbitals with a two-body Jastrow factor and a cusp-enforcing Jastrow term. By writing $\mathbf x := \{\mathbf{r_1}, \mathbf{r_2},..., \mathbf {r_N}\}$, we have
\begin{equation}\label{eq: quantum dot ansatz}
    \psi(\mathbf{x})
    = \det\!\left[\tilde\varphi^{\,\mathrm{BF}}_{k}\!\left(\mathbf{r}_i;
      \{\mathbf{r}_j\}_{j\neq i}\right)\right]
      \, e^{J(\mathbf{x}) + J_{\mathrm{cusp}}(\mathbf{x})}.
\end{equation}
Both the orbitals and the Jastrow factor are expanded in the basis of Eq.~\ref{eq: fock darwin}. The $N$ occupied orbitals are constructed as linear combinations of the basis functions with complex variational parameters $C_{i,j}$,
\begin{equation}
    \tilde\varphi_{k}(\mathbf{r})
    = \sum_{\alpha} C_{\alpha k}\, \varphi_{\alpha}(\mathbf{r}),
    \qquad k = 1, \dots, N.
\end{equation}
On the other hand, the backflow makes each single particle orbital depend on the positions of all electrons, through the transformation:
\begin{equation}
    \tilde\varphi^{\,\mathrm{BF}}_{k}\!\left(\mathbf{r}_i;
      \{\mathbf{r}_j\}_{j\neq i}\right)
    = \tilde\varphi_{k}(\mathbf{r}_i)
      \left[ 1 + \sum_{\alpha,\beta} \varphi_{\alpha}(\mathbf{r}_i)\,
      B^{(k)}_{\alpha\beta}\, \frac{1}{N-1}\sum_{j\neq i}
      \varphi_{\beta}(\mathbf{r}_j) \right],
\end{equation}
with $B^{(k)}_{i,j}$ complex variational parameters.
As the environment of electron $i$ enters only through a permutation-invariant average over the other electrons, exchanging two electrons still exchanges two rows of the matrix, so that the determinant stays antisymmetric. Similarly, the two-body Jastrow factor is constructed as:
\begin{equation}
    J(\mathbf{x}) = \frac{1}{2}\sum_{i \neq j} \sum_{\alpha,\beta}
        \varphi_{\alpha}(\mathbf{r}_i)\, A_{\alpha\beta}\,
        \varphi_{\beta}(\mathbf{r}_j)
\end{equation}
where $A$ is a symmetric matrix of variational parameters. Finally, $J_{\mathrm{cusp}}$ enforces Kato's cusp condition~\cite{pack1966cusp}, which smooth basis expansions cannot
reproduce. In two dimensions this condition fixes the slope of the Jastrow factor at electron coalescence:
\begin{equation}
    J_{\mathrm{cusp}}(\mathbf{x}) = -\frac{1}{3}\sum_{i<j} \frac{\, b^2}{b +  |\mathbf{r}_i - \mathbf{r}_j|},
\end{equation}
where $b$ is a single complex variational parameter. \\

Once the model is defined, we pick as proposer RWM and we pair it with the $\texttt{MCSamplerCont}$. Then the workflow remains identical to the discrete case.
\begin{lstlisting}[style=pythonstyle]
E_c_det, E_c_jastrow, E_c_bf = 6, 4, 4
n_samples = 2**11

net = SlaterJastrowBackflow(
    n_particles=N,
    orbitals=HOOrbitals(E_c_det, omega),
    jastrow=OrbitalJastrow(HOOrbitals(E_c_jastrow, omega)),
    cusp=ElectronCusp(),
    backflow=OrbitalBackflow(HOOrbitals(E_c_bf, omega)),
)
psi = jVMC.vqs.NQS(net, sample_shape=n_dim * N, batch_size=n_samples, seed=123)
proposer = jVMC.propose.RWM(geometry)
sampler = jVMC.sampler.MCSamplerCont(
    psi, proposer, key=123, num_chains=n_samples // 4, num_samples=n_samples,
    sweep_steps=2 * N, thermalization_sweeps=5
)

print('Number of parameters:', psi.num_parameters)
print('Holomorphic:', psi.holomorphic)
\end{lstlisting}
\begin{lstlisting}[style=outputstyle]
Number of parameters: 827
Holomorphic: True
\end{lstlisting}

Figure~\ref{fig: quantum dot} shows the variational energy mean and variance during the optimization compared with the FCI reference. The final VMC energy agrees with the FCI value within a relative error of $2\cdot10^{-4}$. Note that FCI is variational only within its truncated basis and converges slowly for this system because of the electron-electron cusp. \\

The complete example, with the used architecture and the FCI code, can be found in the folder of examples of jVMC.

\subsection{Unitary Dynamics}
\begin{figure}[t!]
  \centering
  \includegraphics[width=\linewidth]{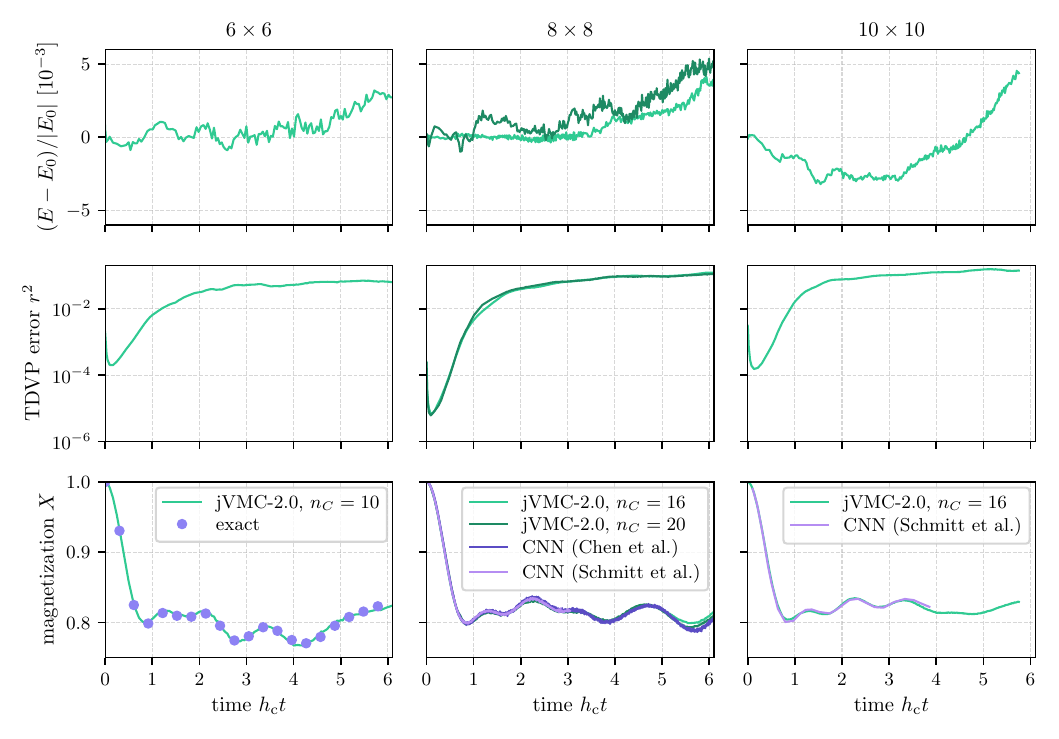}
  \caption{Time evolution after a quench to the critical point of the two-dimensional transverse-field Ising model on $6\times6$ (left), $8\times8$ (middle) and $10\times10$ (right) lattices. Top: energy relative to the initial energy $E_0$, exhibiting statistical fluctuations. Middle: TDVP error $r^2$, rising rapidly after $h_ct\approx1$. Bottom: magnetization $X$, compared with the exact reference for $6\times6$ and with the NQS results of Chen et al.~\cite{chenConvolutionalTransformerWave2025a} and Schmitt et al.~\cite{schmitt2020quantum} for the larger lattices.}
  \label{fig:dynamics_benchmark}
\end{figure}

As a benchmark for real-time evolution with the TDVP, we consider the two-dimensional transverse-field Ising model on an $L\times L$ square lattice with periodic boundary conditions,
\begin{equation}
    \hat H = -J\sum_{\langle i,j \rangle} \hat\sigma^z_i \hat\sigma^z_j - h \sum_{i=1}^{N} \hat\sigma^x_i ,
    \label{eq: tfim 2d}
\end{equation}
where $J>0$, $\langle i,j\rangle$ runs over nearest-neighbour pairs and $N=L^2$. In the ground state, the model undergoes a quantum phase transition between a ferromagnetic and a paramagnetic phase at $h_c/J = 3.04438$~\cite{bloteClusterMonteCarlo2002}. We prepare the system in the $x$-polarized product state $\ket{\Psi_0}=\ket{\rightarrow\rightarrow\cdots\rightarrow}$, which is the ground state of $\hat H$ for $h\to\infty$, and suddenly quench the field to the critical value $h=h_c$. Entanglement grows quickly after such a quench, which makes the dynamics hard for tensor-network methods in two dimensions. It therefore serves as a standard test for NQS~\cite{schmitt2020quantum, chenConvolutionalTransformerWave2025a}. We follow the evolution up to $h_c t = 6$ on lattices of size $6\times6$, $8\times8$ and $10\times10$. For $6\times6$ we compare with exact numerical results, while for the larger lattices we compare with the NQS simulations of Ref.~\cite{schmitt2020quantum}, based on convolutional networks, and of Ref.~\cite{chenConvolutionalTransformerWave2025a}, based on convolutional transformers.

As variational ansatz we use a convolutional residual network (ResNet) with real parameters and a complex output, which encodes the modulus and the phase of the amplitude. It consists of two residual blocks of two convolutional layers each, with $n_C$ filters of size $4\times4$ and GELU activations. Circular padding makes every layer equivariant under lattice translations, and summing the output over the lattice makes the amplitude translation invariant. The global spin-flip symmetry $\prod_i\hat\sigma^x_i$, which commutes with $\hat H$ and leaves $\ket{\Psi_0}$ invariant, is imposed by projection with the \texttt{spin\_flip\_symmetry} projector (Sec.~\ref{sec: sym net}). Since the filters do not depend on $L$, the number of parameters is set by $n_C$ alone: $4\,990$, $12\,592$ and $19\,580$ for $n_C=10$, $16$ and $20$. We use $n_C=10$ for $6\times6$, $n_C=16$ and $20$ for $8\times8$, and $n_C=16$ for $10\times10$.

Expectation values are estimated with $N_\mathrm{MC}$ Markov-chain Monte Carlo samples from 100 chains with single spin-flip updates and 25 thermalization sweeps, where $N_\mathrm{MC}=8\times10^4$ for $6\times6$, and $N_\mathrm{MC}=3\times10^5$ for the larger lattices.

Because the network has real parameters but a complex output, it is not holomorphic, and we integrate the McLachlan form of the TDVP, Eq.~\eqref{eq: tdvp mclachlan}. As previously discussed, diagonal regularization biases the dynamics. We therefore solve the linear system with a plain pseudo-inverse of the quantum geometric tensor, with relative eigenvalue cutoff $10^{-8}$, no diagonal shift and no signal-to-noise regularization. The equation of motion is integrated with an adaptive stepper whose error is measured in the norm induced by the quantum geometric tensor (Sec.~\ref{sec: stepper}), starting from $J\,\delta t = 10^{-3}$ with a maximal step of $J\,\delta t = 0.05$. In particular, we use a second-order adaptive Heun scheme with tolerance $10^{-4}$ for $6\times6$, and RK45 with absolute tolerance $10^{-6}$ for the larger lattices.

The three rows of Fig.~\ref{fig:dynamics_benchmark} show the energy, the TDVP error $r^2$ and the mean magnetization $X=\frac1N\sum_i\langle\sigma^x_i\rangle$, each column corresponding to one system size.
The energy is plotted relative to its initial value $E_0=-h_c N$, which is conserved under the exact dynamics, so that the deviations show the drift and the statistical fluctuations of the simulation.

To monitor the quality of the time evolution, we use the TDVP error $r^2$ introduced in Ref.~\cite{schmitt2020quantum}, quantifying the squared distance between the exact infinitesimal time step and its projection onto the variational manifold, normalized by the norm of the exact step,
\begin{equation}
    r^2 = 1 + \frac{\dot\theta^\dagger S\,\dot\theta - 2\,\mathrm{Re}[\,\dot\theta^\dagger(-iF)]}{\mathrm{Var}(H)},
\end{equation}
where $\dot\theta$ is the TDVP update, $S$ the quantum geometric tensor and $F$ the force vector, so that $r^2=0$ corresponds to an exactly representable step and $r^2=1$ to no overlap at all.
Since all quantities entering the TDVP error have already been computed for the TDVP step itself, $r^2$ is obtained from the same Monte Carlo samples at negligible extra cost; the TDVP optimizer therefore computes it by default and records it in the per-step metadata.
Note however that $r^2$ measures the error of a single step and carries the statistical noise of the estimators; as such, it does not necessarily accumulate the error along the trajectory, but can be understood as its upper bound.

In our quench example, the TDVP error rises rapidly after $h_c t\approx1$ and saturates at a similar level for all system sizes, while the magnetization agrees well with the exact reference for $6\times6$ and with the NQS studies of Refs.~\cite{chenConvolutionalTransformerWave2025a, schmitt2020quantum} for $8\times8$ and $10\times10$, with the two network widths giving nearly indistinguishable results for $8\times8$.
\section{Conclusion}\label{sec: conclusion}

We have presented jVMC 2.0, a redesigned version of the jVMC library for variational Monte Carlo with neural quantum states. The library is organized around a small number of core modules (variational states, samplers, operators, objective functions, optimizers, steppers and solvers), each defined by a minimal abstract interface, so that new components can be combined freely with the existing ones. Building on JAX's sharding mechanism, samples, network evaluations, Jacobians and Jacobian products, such as the QGT and the tangent kernel, are distributed across multiple GPUs and nodes, and the linear systems at the core of SR, MinSR and the TDVP can be solved with distributed, memory-efficient linear algebra thanks to JAXMg.
The support of simulations with mixed floating point precision enables most efficient exploitation of available GPU capacities.
Besides lattice models of spins and fermions, the library now covers particles in continuous space, symmetry projections, autoregressive and transformer architectures, and mixed-precision arithmetic. Our benchmarks show close to linear scaling with the number of GPUs and performance competitive with existing libraries. The examples, ranging from the frustrated $J_1$--$J_2$ model and a two-dimensional quantum dot to quench dynamics in the two-dimensional transverse-field Ising model, illustrate how these components are used together in realistic applications.

The aim of jVMC 2.0 is not to provide every variant of VMC with neural quantum states ready-made. Instead, it offers a solid and performant foundation on which new ideas can be implemented and tested with little effort, and scaled to many devices without additional work. A new network, objective function or optimizer only has to implement the corresponding interface, and it then inherits the distribution across devices, batching and memory management of the rest of the library. The projected time evolution in the examples of the repository is a case in point: it is assembled entirely from existing building blocks, namely the infidelity objective function and the ground-state optimizers.

As for the previous version, the library will continue to be maintained and developed: we will keep it up to date with the evolving JAX ecosystem, review contributions from the community, and add new features over time. Planned additions include restoring the support for positive operator-valued measures (POVMs), which enabled the simulation of open quantum systems in the first version of jVMC; a dedicated optimizer for projected tVMC (ptVMC); an implementation of t-NQS \cite{Walle_2025}; and variants of SR and MinSR, such as SPRING~\cite{goldshlager2024kaczmarz}.
\section*{Code and Data Availability}

The \texttt{jVMC} library is open source (MIT license) and available at
\url{https://github.com/CoQuS/jvmc}.
All scripts and data needed to reproduce the results presented here are available at \url{https://github.com/msolinas28/jvmc-2.0-data}.

\section*{Acknowledgements}
This work was supported via the Helmholtz Initiative and Networking Fund, grant no. VH-NG-1711, the DFG Research Unit FOR5919 “Machine Learning for Complex Quantum States”, grant no. 550495627, and the DFG CRC 1277. Some of the computations were performed on the ADA HPC cluster at the Max Planck Computing and Data Facility. The authors gratefully acknowledge the scientific support and HPC resources provided by the Erlangen National High Performance Computing Center (NHR@FAU) of the Friedrich-Alexander-Universität Erlangen-Nürnberg (FAU) under the NHR project v166da. NHR funding is provided by federal and Bavarian state authorities. The authors thank the Gauss Centre for Supercomputing e.V. (GCS) for providing computing time on the Supercomputer JUPITER at the Jülich Supercomputing Centre. JUPITER is supported by the EuroHPC JU and GCS through funding by the European Commission, the German Federal Ministry of Research, Technology and Space, and the Ministry of Culture and Science of the State of North Rhine-Westphalia.


\begin{appendix}
\numberwithin{equation}{section}
\section{Gradient of the Objective Function}\label{app: grad loss fn}

Here we provide the derivation for Eq.~\ref{eq: grad from force}. We start from the general loss function
\begin{equation}
    \mathcal{L}(\theta)=\sum_i p_i(\theta)\,\ell_i(\theta)\in\mathbb R,
    \qquad
    p_i(\theta)=\frac{|\psi_\theta(s_i)|^2}{\sum_j|\psi_\theta(s_j)|^2},
\end{equation}
where the sum runs over all configurations $s_i$ and $\theta$ are real parameters (in jVMC the
parameters are always treated as real). Here $\ell_i(\theta)=\ell_\theta(s_i)$ is a general,
in principle complex-valued, differentiable function of $\theta$ whose expectation value over $p$
is real for all $\theta$. We further define the logarithmic derivatives
\begin{equation}
    \Gamma_{i,k}=\partial_{\theta_k}\log\psi_\theta(s_i),
\end{equation}
which are complex in general.

Using $|\psi_\theta(s_i)|^2=\exp\left(2\,\text{Re}[\log\psi_\theta(s_i)]\right)$ and the fact that, for real $\theta$, $\partial_k\text{Re}[\,\cdot\,]=\text{Re}[\partial_k\,\cdot\,]$, we have
$\partial_k\log|\psi_\theta(s_i)|^2=2\,\text{Re}(\Gamma_{i,k})$. The normalization then gives $\partial_k\log\sum_j|\psi_\theta(s_j)|^2=2\,\expval[p]{\text{Re}(\Gamma_k)}$, so that
\begin{equation}
    \partial_k p_i = 2p_i\left(\text{Re}(\Gamma_{i,k})-\expval[p]{\text{Re}(\Gamma_k)}\right).
\end{equation}
The derivative of the loss therefore reads
\begin{equation}
    \partial_k\mathcal L
    = 2\,\expval[p]{\left(\text{Re}(\Gamma_k)-\expval[p]{\text{Re}(\Gamma_k)}\right)\ell(\theta)}
    + \expval[p]{\partial_k\ell(\theta)}.
\end{equation}
Since $\expval[p]{\ell(\theta)}$ is a constant with respect to the sampling and
$\expval[p]{\text{Re}(\Gamma_k)-\expval[p]{\text{Re}(\Gamma_k)}}=0$, we can subtract
$\expval[p]{\ell(\theta)}$ from $\ell(\theta)$ in the first term, which turns it into a covariance:
\begin{equation}
    \partial_k\mathcal L = 2\,\text{Cov}_{p}\left(\text{Re}(\Gamma_k),\ell(\theta)\right)
    + \expval[p]{\partial_k\ell(\theta)}.
\end{equation}

Because $\text{Re}(\Gamma_k)$ is real, the right-hand side splits term by term into a real and an imaginary part. The imaginary part is
\begin{equation}
    2\,\text{Cov}_p\left(\text{Re}(\Gamma_k),\text{Im}(\ell)\right)+\expval[p]{\partial_k\text{Im}(\ell)}
=\partial_k\,\text{Im}\,\mathcal L,
\end{equation}
which vanishes identically because $\mathcal L(\theta)$ is a function of with real domain and image. Note that its two contributions cancel each other but are not individually zero. We are thus left with
\begin{equation}
    \partial_k\mathcal L = 2\,\text{Cov}_{p}\left(\text{Re}(\Gamma_k),\text{Re}(\ell)\right)
    + \expval[p]{\partial_k\text{Re}(\ell)}.
\end{equation}

To connect to the standard VMC formulation, we use
\begin{align}
    \text{Re}\,\text{Cov}_p\left(\Gamma_k^*,\ell\right)
    &= \text{Cov}_p\left(\text{Re}(\Gamma_k),\text{Re}(\ell)\right)
    + \text{Cov}_p\left(\text{Im}(\Gamma_k),\text{Im}(\ell)\right)=\\&=\text{Cov}_p\left(\text{Re}(\Gamma_k),\text{Re}(\ell)\right)+\expval[p]{\text{Im}(\Gamma_k)\,\text{Im}(\ell)},
\end{align}
where in the second step we used $\text{Im}\expval[p]{\ell}=\expval[p]{\text{Im}(\ell)}=0$. The gradient can then be written as
\begin{equation}
    \partial_k\mathcal L = 2\,\text{Re}\left(F_k\right),
\end{equation}
with the force vector
\begin{equation}
    F_k = \text{Cov}_p\left(\Gamma_k^*,\ell\right) + \frac12 R_k,
    \qquad
    R_k = \expval[p]{\partial_k\text{Re}(\ell)}
    - 2\,\expval[p]{\text{Im}(\Gamma_k)\,\text{Im}(\ell)} \in \mathbb R.
\end{equation}

As a consistency check, consider the case where $\ell$ is the local estimator of a Hermitian
operator $\hat O$, i.e.\ $\ell(s)=(\hat O\psi_\theta)(s)/\psi_\theta(s)$. Then
\begin{equation}
    \partial_k\ell(s)=\frac{(\hat O\,\partial_k\psi_\theta)(s)}{\psi_\theta(s)}-\ell(s)\,\Gamma_k(s).
\end{equation}
We can use the Hermiticity of $\hat O$ to rewrite the first term as
\begin{align}
    \expval[p]{\frac{\hat O\,\partial_k\psi_\theta}{\psi_\theta}}
    &=\frac{\langle\psi_\theta|\hat O|\partial_k\psi_\theta\rangle}{\langle\psi_\theta|\psi_\theta\rangle}
    =\frac{\langle\hat O\psi_\theta|\partial_k\psi_\theta\rangle}{\langle\psi_\theta|\psi_\theta\rangle}
    \\&=\sum_s p(s)\left(\frac{(\hat O\psi_\theta)(s)}{\psi_\theta(s)}\right)^{\!*}
    \frac{\partial_k\psi_\theta(s)}{\psi_\theta(s)}
    =\expval[p]{\ell^*\,\Gamma_k},
\end{align}
so that
\begin{equation}
    \expval[p]{\partial_k\ell}=\expval[p]{(\ell^*-\ell)\,\Gamma_k}
    =-2i\,\expval[p]{\text{Im}(\ell)\,\Gamma_k}.
\end{equation}

Taking the real part yields
$\expval[p]{\partial_k\text{Re}(\ell)}=2\,\expval[p]{\text{Im}(\Gamma_k)\,\text{Im}(\ell)}$,
hence $R_k=0$, and we recover the standard VMC force $F_k=\text{Cov}_p\left(\Gamma_k^*,\ell\right)$. \\

On the other hand, if the operator depends explicitly on the parameters, $\hat O=\hat O(\theta)$ Hermitian for all $\theta$, the derivative of the local loss acquires the additional term
$(\partial_{\theta_k}\hat O)_\text{loc}(s)=\sum_{s'}\partial_{\theta_k}O_{ss'}(\theta)\,\psi_\theta(s')/\psi_\theta(s)$.
The remaining contributions cancel against the second term of $R_k$ as before, so that
$R_k=\text{Re}\,\expval[p]{(\partial_{\theta_k}\hat O)_\text{loc}}=\expval[p]{(\partial_{\theta_k}\hat O)_\text{loc}}$,
where the last equality holds because $\partial_{\theta_k}\hat O$ is Hermitian.

\section{Memory-Efficient Distributed Jacobian Products}\label{app: distributed Jacobian products}
\begin{figure}[t]
    \centering
    \includegraphics[width=0.8\linewidth]{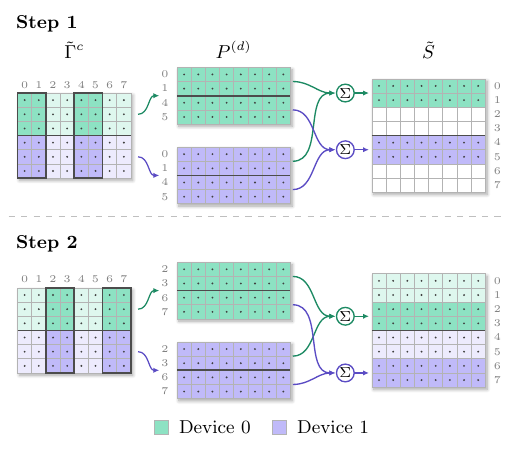}
    \caption{Construction of $\tilde S$ for $N_d=2$ devices, $N_s=6$ samples and $\tilde N_p=8$ padded parameters, so that each device holds a slab of $4$ rows of $\tilde S$, split into blocks of $B=2$ rows. In each step, every device multiplies the selected columns of its local rows of $\tilde\Gamma^c$ (outlined) with all of its local rows, obtaining the local product $P^{(d)}$, whose rows are labeled by their index in $\tilde S$. The local products are then summed over devices ($\Sigma$) and scattered, so that the upper half of every $P^{(d)}$ contributes to the rows of device $0$ and the lower half to the rows of device $1$, filling one block of each slab.}
    \label{fig: gqt construction example}
\end{figure}

In this section we outline how the QGT is built so that it remains distributed across the $N_d$ devices throughout the whole construction.

The main building block is the Jacobian of the log-amplitude,
\begin{equation}
    \Gamma_{i,k} = \frac{\partial}{\partial\theta_k}\log\psi_\theta(s_i), \qquad \text{with shape } (N_s, N_p).
\end{equation}
We can then define the centered Jacobian
\begin{equation}
    \Gamma^c_{i,k} = \frac{1}{\sqrt{N_s}}(\Gamma_{i,k}-\expval{\Gamma_k}),
\end{equation}
so that the QGT reads
\begin{equation}\label{eq:qgt}
    S = \left(\Gamma^c\right)^\dagger\Gamma^c, \qquad \text{with shape } (N_p, N_p).
\end{equation}

The Jacobian returned by \texttt{NQS.gradients()} is distributed over the sample axis: device $d$ holds a shard $\Gamma^{c,(d)}$ of $N_s/N_d$ rows, so that Eq.~\eqref{eq:qgt} is a sum of local products, 
\begin{equation}
    S = \sum_d (\Gamma^{c,(d)})^\dagger\Gamma^{c,(d)}
\end{equation}

We therefore need a way of obtaining the QGT with two constraints in mind: (i) $S$ has to be distributed over its first axis, with each device holding a contiguous slab of $N_p/N_d$ rows, and
(ii) no device has to ever hold an array larger than $(N_p/N_d)\times N_p$, i.e.\ the size of its own slab.
The naive approach, in which each device computes its full local product and the results are then summed and scattered, violates (ii), since every local product has shape $(N_p, N_p)$.

To satisfy both requirements, we first pad the Jacobian with zero columns so that the number of parameters is divisible by $N_d^2$. We call the padded matrix $\tilde\Gamma^c$, with shape $(N_s, \tilde N_p)$. We then split the slab of each device into $N_d$ blocks of $B=\tilde N_p/N_d^2$ rows, and denote by $\mathcal{R}_{d,j}$ the rows of the $j$-th block of device $d$:
\begin{equation}
    \mathcal{R}_{d,j} = \left\{\, d\,\frac{\tilde N_p}{N_d} + jB + b \;:\; b=0,\dots,B-1 \right\}.
\end{equation}
The algorithm proceeds in $N_d$ sequential steps, where the $j$-th steps reads:
\begin{enumerate}
    \item Select the columns $\mathcal{C}_j = \mathcal{R}_{0,j}\cup\mathcal{R}_{1,j}\cup\dots\cup\mathcal{R}_{N_d-1,j}$ of $\tilde\Gamma^c$, i.e.\ the $j$-th block of every slab, ordered by device. These are $\tilde N_p/N_d$ columns in total.
    \item Each device $d$ multiplies its selected columns with its whole shard,
    \begin{equation}
        P^{(d)}_j = \left(\tilde\Gamma^{c,(d)}_{:,\mathcal{C}_j}\right)^\dagger\tilde\Gamma^{c,(d)}, \qquad \text{with shape } (\tilde N_p/N_d, \tilde N_p),
    \end{equation}
    which is the contribution of its local samples to the rows $\mathcal{C}_j$ of $\tilde S$.
    \item The matrices $P^{(d)}_j$ are summed over devices and scattered along their first axis, so that device $d$ receives the $d$-th group of $B$ rows of $\sum_{d'}P^{(d')}_j$. Because of the ordering chosen in point (1), these are exactly the rows $\mathcal{R}_{d,j}$ of $\tilde S$, i.e.\ the $j$-th block of the slab of device $d$, which is written in place. No further communication between devices is needed.
\end{enumerate}
After $N_d$ steps every device holds its full slab of $\tilde S$, and the largest array ever held by a device is $P^{(d)}_j$, which has the same shape as the slab itself. Finally, since the product in point (2) is a sum over samples, it can equivalently be accumulated over batches of samples to further reduce the memory footprint, without changing the rest of the algorithm, which is exactly what happens under the hood when the Jacobian is given as $\texttt{LazySampledObs}$. We give an example of this algorithm in Fig.~\ref{fig: gqt construction example}. \\

The tangent kernel used in MinSR, $T=\tilde\Gamma^c(\tilde\Gamma^c)^\dagger$ with shape $(N_s,N_s)$, is constructed with the same approach and one additional step. Since the product now sums over the parameters rather than over the samples, the Jacobian is first redistributed with an all-to-all communication, so that each device holds all $N_s$ samples but only $\tilde N_p/N_d$ parameters, with the parameters padded to a multiple of $N_d$. The local product on each device is then a partial sum over its own parameters, and the algorithm described above applies with the roles of samples and parameters exchanged.
\end{appendix}

\bibliography{bib.bib}

@software{flax2020github,
  author = {Jonathan Heek and Anselm Levskaya and Avital Oliver and Marvin Ritter and Bertrand Rondepierre and Andreas Steiner and Marc van {Z}ee},
  title = {{F}lax: A neural network library and ecosystem for {JAX}},
  url = {http://github.com/google/flax},
  version = {0.12.8},
  year = {2024},
}

@software{jax2018github,
  author = {James Bradbury and Roy Frostig and Peter Hawkins and Matthew James Johnson and Yash Katariya and Chris Leary and Dougal Maclaurin and George Necula and Adam Paszke and Jake Vander{P}las and Skye Wanderman-{M}ilne and Qiao Zhang},
  title = {{JAX}: composable transformations of {P}ython+{N}um{P}y programs},
  url = {http://github.com/jax-ml/jax},
  version = {0.3.13},
  year = {2018},
}

@inproceedings{
solinas2026neural,
title={Neural Quantum States in Mixed Precision},
author={Massimo Solinas and Agnes Valenti and Nawaf Bou-Rabee and Roeland Wiersema},
booktitle={Forty-third International Conference on Machine Learning},
year={2026},
url={https://openreview.net/forum?id=aDLJNoseDH}
}

@article{vats2021revisiting,
  title={Revisiting the gelman--rubin diagnostic},
  author={Vats, Dootika and Knudson, Christina},
  journal={Statistical Science},
  volume={36},
  number={4},
  pages={518--529},
  year={2021},
  publisher={JSTOR}
}

@article{peng2023rwkv,
  title={Rwkv: Reinventing rnns for the transformer era},
  author={Peng, Bo and Alcaide, Eric and Anthony, Quentin and Albalak, Alon and Arcadinho, Samuel and Biderman, Stella and Cao, Huanqi and Cheng, Xin and Chung, Michael and Grella, Matteo and others},
  journal={arXiv preprint arXiv:2305.13048},
  year={2023}
}

@article{rigo2025operator,
  title={Operator Lanczos Approach enabling Neural Quantum States as Real-Frequency Impurity Solvers},
  author={Rigo, Jonas B and Schmitt, Markus},
  journal={arXiv preprint arXiv:2512.08624},
  year={2025}
}

@article{wiersema2026numerical,
  title={Numerical simulation of D-Wave's quantum advantage experiment with time-dependent variational Monte Carlo},
  author={Wiersema, Roeland},
  journal={arXiv preprint arXiv:2609.01719},
  year={2026}
}

@article{kingma2014adam,
  title={Adam: A method for stochastic optimization},
  author={Kingma, Diederik P and Ba, Jimmy},
  journal={arXiv preprint arXiv:1412.6980},
  year={2014}
}

@book{becca2017quantum,
  title={Quantum Monte Carlo approaches for correlated systems},
  author={Becca, Federico and Sorella, Sandro},
  year={2017},
  publisher={Cambridge University Press}
}

@article{sorella1998green,
  title={Green function Monte Carlo with stochastic reconfiguration},
  author={Sorella, Sandro},
  journal={Physical review letters},
  volume={80},
  number={20},
  pages={4558},
  year={1998},
  publisher={APS}
}

@article{chen2024empowering,
  title={Empowering deep neural quantum states through efficient optimization},
  author={Chen, Ao and Heyl, Markus},
  journal={Nature Physics},
  volume={20},
  number={9},
  pages={1476--1481},
  year={2024},
  publisher={Nature Publishing Group UK London}
}

@article{rende2024simple,
  title={A simple linear algebra identity to optimize large-scale neural network quantum states},
  author={Rende, Riccardo and Viteritti, Luciano Loris and Bardone, Lorenzo and Becca, Federico and Goldt, Sebastian},
  journal={Communications Physics},
  volume={7},
  number={1},
  pages={260},
  year={2024},
  publisher={Nature Publishing Group UK London}
}

@inproceedings{amari1998natural,
  title={Why natural gradient?},
  author={Amari, Shun-Ichi and Douglas, Scott C},
  booktitle={Proceedings of the 1998 IEEE International Conference on Acoustics, Speech and Signal Processing, ICASSP'98 (Cat. No. 98CH36181)},
  volume={2},
  pages={1213--1216},
  year={1998},
  organization={IEEE}
}

@article{hackl2020geometry,
  title={Geometry of variational methods: dynamics of closed quantum systems},
  author={Hackl, Lucas and Guaita, Tommaso and Shi, Tao and Haegeman, Jutho and Demler, Eugene and Cirac, J Ignacio},
  journal={SciPost Physics},
  volume={9},
  number={4},
  pages={048},
  year={2020}
}

@article{medvidovic2023variational,
  title={Variational quantum dynamics of two-dimensional rotor models},
  author={Medvidovi{\'c}, Matija and Sels, Dries},
  journal={PRX Quantum},
  volume={4},
  number={4},
  pages={040302},
  year={2023},
  publisher={APS}
}

@article{schmitt2020quantum,
  title={Quantum many-body dynamics in two dimensions with artificial neural networks},
  author={Schmitt, Markus and Heyl, Markus},
  journal={Physical Review Letters},
  volume={125},
  number={10},
  pages={100503},
  year={2020},
  publisher={APS}
}

@article{wiersema2026jaxmg,
  title={JAXMg: A multi-GPU linear solver in JAX},
  author={Wiersema, Roeland},
  journal={arXiv preprint arXiv:2601.14466},
  year={2026}
}

@article{rossky1978brownian,
  title={Brownian dynamics as smart Monte Carlo simulation},
  author={Rossky, Peter J and Doll, Jimmie D and Friedman, Harold L},
  journal={The Journal of Chemical Physics},
  volume={69},
  number={10},
  pages={4628--4633},
  year={1978},
  publisher={American Institute of Physics}
}

@article{metropolis1953equation,
  title={Equation of state calculations by fast computing machines},
  author={Metropolis, Nicholas and Rosenbluth, Arianna W and Rosenbluth, Marshall N and Teller, Augusta H and Teller, Edward},
  journal={The journal of chemical physics},
  volume={21},
  number={6},
  pages={1087--1092},
  year={1953},
  publisher={American Institute of Physics}
}

@misc{chenConvolutionalTransformerWave2025a,
  title = {Convolutional Transformer Wave Functions},
  author = {Chen, Ao and Naik, Vighnesh Dattatraya and Heyl, Markus},
  year = 2025,
  month = mar,
  number = {arXiv:2503.10462},
  eprint = {2503.10462},
  primaryclass = {cond-mat.dis-nn},
  publisher = {arXiv},
  doi = {10.48550/arXiv.2503.10462},
  urldate = {2026-08-12},
  archiveprefix = {arXiv},
}

@misc{krinitsinTimedependentVariationalMonte2026,
  title = {Time-Dependent Variational {{Monte Carlo}} without Bias},
  author = {Krinitsin, Wladislaw and Schmitt, Markus},
  year = 2026,
  month = may,
  journal = {arXiv},
  eprint = {2605.03930},
  primaryclass = {quant-ph)},
  archiveprefix = {arXiv},
}

@article{sinibaldiUnbiasingTimedependentVariational2023,
  title = {Unbiasing Time-Dependent {{Variational Monte Carlo}} by Projected Quantum Evolution},
  author = {Sinibaldi, Alessandro and Giuliani, Clemens and Carleo, Giuseppe and Vicentini, Filippo},
  year = 2023,
  month = oct,
  journal = {Quantum},
  volume = {7},
  pages = {1131},
  issn = {2521-327X},
  doi = {10.22331/q-2023-10-10-1131},
  urldate = {2026-09-07},
  langid = {english},
}

@article{gravina2025neural,
  title={Neural projected quantum dynamics: a systematic study},
  author={Gravina, Luca and Savona, Vincenzo and Vicentini, Filippo},
  journal={Quantum},
  volume={9},
  pages={1803},
  year={2025},
  publisher={Verein zur F{\"o}rderung des Open Access Publizierens in den Quantenwissenschaften}
}

@article{anisimovas1998energy,
  title={Energy spectra of few-electron quantum dots},
  author={Anisimovas, E and Matulis, A},
  journal={Journal of Physics: Condensed Matter},
  volume={10},
  number={3},
  pages={601--615},
  year={1998}
}

@article{nys2024ab,
  title={Ab-initio variational wave functions for the time-dependent many-electron Schr{\"o}dinger equation},
  author={Nys, Jannes and Pescia, Gabriel and Sinibaldi, Alessandro and Carleo, Giuseppe},
  journal={Nature communications},
  volume={15},
  number={1},
  pages={9404},
  year={2024},
  publisher={Nature Publishing Group UK London}
}

@article{pack1966cusp,
  title={Cusp conditions for molecular wavefunctions},
  author={Pack, Russell T and Brown, W Byers},
  journal={The Journal of Chemical Physics},
  volume={45},
  number={2},
  pages={556--559},
  year={1966},
  publisher={American Institute of Physics}
}

@article{vicentini2021netket,
  title={NetKet 3: Machine learning toolbox for many-body quantum systems},
  author={Vicentini, Filippo and Hofmann, Damian and Szab{\'o}, Attila and Wu, Dian and Roth, Christopher and Giuliani, Clemens and Pescia, Gabriel and Nys, Jannes and Vargas-Calder{\'o}n, Vladimir and Astrakhantsev, Nikita and others},
  journal={arXiv preprint arXiv:2112.10526},
  year={2021}
}

@article{moss2025double,
  title={Double descent: When do neural quantum states generalize?},
  author={Moss, M Schuyler and Orfi, Alev and Roth, Christopher and Sengupta, Anirvan M and Georges, Antoine and Sels, Dries and Dawid, Anna and Valenti, Agnes},
  journal={arXiv preprint arXiv:2508.00068},
  year={2025}
}

@article{king2025beyond,
  title={Beyond-classical computation in quantum simulation},
  author={King, Andrew D and Nocera, Alberto and Rams, Marek M and Dziarmaga, Jacek and Wiersema, Roeland and Bernoudy, William and Raymond, Jack and Kaushal, Nitin and Heinsdorf, Niclas and Harris, Richard and others},
  journal={Science},
  volume={388},
  number={6743},
  pages={199--204},
  year={2025},
  publisher={American Association for the Advancement of Science}
}

@article{carleo2017solving,
  title={Solving the quantum many-body problem with artificial neural networks},
  author={Carleo, Giuseppe and Troyer, Matthias},
  journal={Science},
  volume={355},
  number={6325},
  pages={602--606},
  year={2017},
  publisher={American Association for the Advancement of Science}
}

@article{rigo2026neural,
  title={Neural quantum states in condensed matter: advances, best practices, and prospects},
  author={Rigo, Jonas B and Wurst, Bj{\"o}rn J and Nutakki, Rajah and Schmitt, Markus and Kennes, Dante},
  journal={arXiv preprint arXiv:2608.21291},
  year={2026}
}

@article{medvidovic2024neural,
  title={Neural-network quantum states for many-body physics},
  author={Medvidovi{\'c}, Matija and Moreno, Javier Robledo},
  journal={The European Physical Journal Plus},
  volume={139},
  number={7},
  pages={1--26},
  year={2024},
  publisher={Springer}
}

@article{choo2019two,
  title   = {Two-dimensional frustrated ${J}_1-{J}_2$ model studied with neural network quantum states},
  author  = {Choo, Kenny and Neupert, Titus and Carleo, Giuseppe},
  journal = {Physical Review B},
  volume  = {100},
  pages   = {125124},
  year    = {2019},
  doi     = {10.1103/PhysRevB.100.125124}
}

@article{levine2019quantum,
  title={Quantum entanglement in deep learning architectures},
  author={Levine, Yoav and Sharir, Or and Cohen, Nadav and Shashua, Amnon},
  journal={Physical review letters},
  volume={122},
  number={6},
  pages={065301},
  year={2019},
  publisher={APS}
}

@article{sharir2022neural,
  title={Neural tensor contractions and the expressive power of deep neural quantum states},
  author={Sharir, Or and Shashua, Amnon and Carleo, Giuseppe},
  journal={Physical Review B},
  volume={106},
  number={20},
  pages={205136},
  year={2022},
  publisher={APS}
}

@article{hornik1991approximation,
  title={Approximation capabilities of multilayer feedforward networks},
  author={Hornik, Kurt},
  journal={Neural networks},
  volume={4},
  number={2},
  pages={251--257},
  year={1991},
  publisher={Elsevier}
}

@article{pfau2024accurate,
  title={Accurate computation of quantum excited states with neural networks},
  author={Pfau, David and Axelrod, Simon and Sutterud, Halvard and Von Glehn, Ingrid and Spencer, James S},
  journal={Science},
  volume={385},
  number={6711},
  pages={eadn0137},
  year={2024},
  publisher={American Association for the Advancement of Science}
}

@article{von2022self,
  title={A self-attention ansatz for ab-initio quantum chemistry},
  author={von Glehn, Ingrid and Spencer, James S and Pfau, David},
  journal={arXiv preprint arXiv:2211.13672},
  year={2022}
}

@article{fu2026towards,
  title={Towards stable and accurate electron dynamics via neural network based time-dependent variational Monte Carlo},
  author={Fu, Weizhong and Li, Zhe and Qian, Yubing and Li, Ruichen and Ren, Weiluo and Chen, Ji},
  journal={arXiv preprint arXiv:2606.05850},
  year={2026}
}

@article{chen2026neural,
  title={Neural network--augmented Pfaffian wave-functions for scalable simulations of interacting fermions},
  author={Chen, Ao and Wan, Zhou-Quan and Sengupta, Anirvan and Georges, Antoine and Roth, Christopher},
  journal={Proceedings of the National Academy of Sciences},
  volume={123},
  number={35},
  pages={e2535288123},
  year={2026},
  publisher={National Academy of Sciences}
}

@article{lange2026simulating,
  title   = {Simulating superconductivity in mixed-dimensional $t_\parallel$-$J_\parallel$-$J_\perp$ bilayers with neural quantum states},
  author  = {Lange, Hannah and Chen, Ao and Georges, Antoine and Grusdt, Fabian and Bohrdt, Annabelle and Roth, Christopher},
  journal = {arXiv preprint arXiv:2602.10091},
  year    = {2026}
}

@article{medvidovic2026adiabatic,
  title={Adiabatic transport of neural network quantum states},
  author={Medvidovi{\'c}, Matija and Orfi, Alev and Carrasquilla, Juan and Sels, Dries},
  journal={PRX Intelligence},
  volume={1},
  number={1},
  pages={013020},
  year={2026},
  publisher={APS}
}

@article{sharir2020deep,
  title   = {Deep Autoregressive Models for the Efficient Variational Simulation of Many-Body Quantum Systems},
  author  = {Sharir, Or and Levine, Yoav and Wies, Noam and Carleo, Giuseppe and Shashua, Amnon},
  journal = {Physical Review Letters},
  volume  = {124},
  pages   = {020503},
  year    = {2020},
  doi     = {10.1103/PhysRevLett.124.020503}
}

@article{duque2026one,
  title={One More Time: Revisiting Neural Quantum States from a Reinforcement Learning Perspective},
  author={Duque, Juan Agust{\'\i}n and Heredia, Sergio Garc{\'\i}a and Hernandes, Vinicius and Greplov{\'a}, Eli{\v{s}}ka and Spriggs, Thomas and Courville, Aaron and Dawid, Anna},
  journal={arXiv preprint arXiv:2607.02292},
  year={2026}
}

@article{krinitsin2026geometric,
  title={Geometric inflation of deviations challenges neural quantum states in dynamics of quantum Ising models},
  author={Krinitsin, Wladislaw and Rigo, Jonas B and Abedi, Mohammad and Schmitt, Markus},
  journal={arXiv preprint arXiv:2609.07645},
  year={2026}
}

@article{schmitt2021jvmc,
  title={jVMC: Versatile and performant variational Monte Carlo leveraging automated differentiation and GPU acceleration},
  author={Schmitt, Markus and Reh, Moritz},
  journal={arXiv preprint arXiv:2108.03409},
  year={2021}
}

@incollection{weisse2008exact,
  title={Exact diagonalization techniques},
  author={Wei{\ss}e, Alexander and Fehske, Holger},
  booktitle={Computational many-particle physics},
  pages={529--544},
  year={2008},
  publisher={Springer}
}

@book{szabo2012modern,
  title={Modern quantum chemistry: introduction to advanced electronic structure theory},
  author={Szabo, Attila and Ostlund, Neil S},
  year={2012},
  publisher={Courier Corporation}
}

@article{foulkes2001quantum,
  title={Quantum Monte Carlo simulations of solids},
  author={Foulkes, William MC and Mitas, Lubos and Needs, RJ and Rajagopal, Guna},
  journal={Reviews of Modern Physics},
  volume={73},
  number={1},
  pages={33},
  year={2001},
  publisher={APS}
}

@incollection{assaad2008world,
  title={World-line and determinantal quantum Monte Carlo methods for spins, phonons and electrons},
  author={Assaad, FF and Evertz, HG},
  booktitle={Computational many-particle physics},
  pages={277--356},
  year={2008},
  publisher={Springer}
}

@article{troyer2005computational,
  title={Computational complexity and fundamental limitations to fermionic quantum Monte Carlo simulations},
  author={Troyer, Matthias and Wiese, Uwe-Jens},
  journal={Physical review letters},
  volume={94},
  number={17},
  pages={170201},
  year={2005},
  publisher={APS}
}

@article{white1992density,
  title   = {Density matrix formulation for quantum renormalization groups},
  author  = {White, Steven R.},
  journal = {Physical Review Letters},
  volume  = {69},
  pages   = {2863--2866},
  year    = {1992},
  doi     = {10.1103/PhysRevLett.69.2863}
}

@article{schollwock2011density,
  title   = {The density-matrix renormalization group in the age of matrix product states},
  author  = {Schollw{\"o}ck, Ulrich},
  journal = {Annals of Physics},
  volume  = {326},
  pages   = {96--192},
  year    = {2011},
  doi     = {10.1016/j.aop.2010.09.012}
}

@article{orus2014practical,
  title   = {A practical introduction to tensor networks: Matrix product states and projected entangled pair states},
  author  = {Or{\'u}s, Rom{\'a}n},
  journal = {Annals of Physics},
  volume  = {349},
  pages   = {117--158},
  year    = {2014},
  doi     = {10.1016/j.aop.2014.06.013}
}

@article{eisert2010colloquium,
  title   = {Colloquium: Area laws for the entanglement entropy},
  author  = {Eisert, J. and Cramer, M. and Plenio, M. B.},
  journal = {Reviews of Modern Physics},
  volume  = {82},
  pages   = {277--306},
  year    = {2010},
  doi     = {10.1103/RevModPhys.82.277}
}

@article{metzner1989correlated,
  title   = {Correlated Lattice Fermions in $d=\infty$ Dimensions},
  author  = {Metzner, Walter and Vollhardt, Dieter},
  journal = {Physical Review Letters},
  volume  = {62},
  pages   = {324--327},
  year    = {1989},
  doi     = {10.1103/PhysRevLett.62.324}
}

@article{georges1996dynamical,
  title   = {Dynamical mean-field theory of strongly correlated fermion systems and the limit of infinite dimensions},
  author  = {Georges, Antoine and Kotliar, Gabriel and Krauth, Werner and Rozenberg, Marcelo J.},
  journal = {Reviews of Modern Physics},
  volume  = {68},
  pages   = {13--125},
  year    = {1996},
  doi     = {10.1103/RevModPhys.68.13}
}

@article{kotliar2006electronic,
  title   = {Electronic structure calculations with dynamical mean-field theory},
  author  = {Kotliar, G. and Savrasov, S. Y. and Haule, K. and Oudovenko, V. S. and Parcollet, O. and Marianetti, C. A.},
  journal = {Reviews of Modern Physics},
  volume  = {78},
  pages   = {865--951},
  year    = {2006},
  doi     = {10.1103/RevModPhys.78.865}
}

@article{maier2005quantum,
  title   = {Quantum cluster theories},
  author  = {Maier, Thomas and Jarrell, Mark and Pruschke, Thomas and Hettler, Matthias H.},
  journal = {Reviews of Modern Physics},
  volume  = {77},
  pages   = {1027--1080},
  year    = {2005},
  doi     = {10.1103/RevModPhys.77.1027}
}

@article{deng2017quantum,
  title   = {Quantum Entanglement in Neural Network States},
  author  = {Deng, Dong-Ling and Li, Xiaopeng and Das Sarma, S.},
  journal = {Physical Review X},
  volume  = {7},
  pages   = {021021},
  year    = {2017},
  doi     = {10.1103/PhysRevX.7.021021}
}

@article{hibat2020recurrent,
  title   = {Recurrent neural network wave functions},
  author  = {Hibat-Allah, Mohamed and Ganahl, Martin and Hayward, Lauren E. and Melko, Roger G. and Carrasquilla, Juan},
  journal = {Physical Review Research},
  volume  = {2},
  pages   = {023358},
  year    = {2020},
  doi     = {10.1103/PhysRevResearch.2.023358}
}

@article{lange2024architectures,
  title={From architectures to applications: A review of neural quantum states},
  author={Lange, Hannah and Van de Walle, Anka and Abedinnia, Atiye and Bohrdt, Annabelle},
  journal={Quantum Science and Technology},
  volume={9},
  number={4},
  pages={040501},
  year={2024},
  publisher={IOP Publishing}
}

@article{viteritti2023transformer,
  title   = {Transformer Variational Wave Functions for Frustrated Quantum Spin Systems},
  author  = {Viteritti, Luciano Loris and Rende, Riccardo and Becca, Federico},
  journal = {Physical Review Letters},
  volume  = {130},
  pages   = {236401},
  year    = {2023},
  doi     = {10.1103/PhysRevLett.130.236401}
}

@article{hermann2020deep,
  title   = {Deep-neural-network solution of the electronic {S}chr{\"o}dinger equation},
  author  = {Hermann, Jan and Sch{\"a}tzle, Zeno and No{\'e}, Frank},
  journal = {Nature Chemistry},
  volume  = {12},
  pages   = {891--897},
  year    = {2020},
  doi     = {10.1038/s41557-020-0544-y}
}

@article{pfau2020ab,
  title   = {Ab initio solution of the many-electron {S}chr{\"o}dinger equation with deep neural networks},
  author  = {Pfau, David and Spencer, James S. and Matthews, Alexander G. D. G. and Foulkes, W. M. C.},
  journal = {Physical Review Research},
  volume  = {2},
  pages   = {033429},
  year    = {2020},
  doi     = {10.1103/PhysRevResearch.2.033429}
}

@article{bloteClusterMonteCarlo2002,
  title = {Cluster {{Monte Carlo}} Simulation of the Transverse {{Ising}} Model},
  author = {Bl{\"o}te, Henk W. J. and Deng, Youjin},
  year = 2002,
  month = dec,
  journal = {Physical Review E},
  volume = {66},
  number = {6},
  pages = {066110},
  issn = {1063-651X, 1095-3787},
  doi = {10.1103/PhysRevE.66.066110},
  urldate = {2026-10-21},
  copyright = {http://link.aps.org/licenses/aps-default-license},
  langid = {english}
}

@article{goldshlager2024kaczmarz,
  title={A Kaczmarz-inspired approach to accelerate the optimization of neural network wavefunctions},
  author={Goldshlager, Gil and Abrahamsen, Nilin and Lin, Lin},
  journal={Journal of Computational Physics},
  volume={516},
  pages={113351},
  year={2024},
  publisher={Elsevier}
}

@article{Walle_2025,
doi = {10.1088/2632-2153/ae0f39},
url = {https://doi.org/10.1088/2632-2153/ae0f39},
year = {2025},
month = {oct},
publisher = {IOP Publishing},
volume = {6},
number = {4},
pages = {045011},
author = {Van de Walle, Anka and Schmitt, Markus and Bohrdt, Annabelle},
title = {Many-body dynamics with explicitly time-dependent neural quantum states},
journal = {Machine Learning: Science and Technology}
}

@article{viteritti2025square,
  title   = {Transformer wave function for two dimensional frustrated magnets:
             Emergence of a spin-liquid phase in the Shastry--Sutherland model},
  author  = {Viteritti, Luciano Loris and Rende, Riccardo and Parola, Alberto
             and Goldt, Sebastian and Becca, Federico},
  journal = {Phys. Rev. B},
  volume  = {111},
  pages   = {134411},
  year    = {2025},
  doi     = {10.1103/PhysRevB.111.134411}
}

\end{document}